\documentclass[pdftex,colorlinks=true,linkcolor=Chartreuse3,twocolumn,prd,superscriptaddress,nofootinbib,aps]{revtex4-1}
\usepackage[dvipsnames,svgnames,x11names]{xcolor}
\usepackage{amsmath}
\usepackage{amssymb}
\usepackage{amsfonts}
\usepackage{comment}
\usepackage{graphicx}
\usepackage[colorlinks=true,citecolor=Chartreuse3,urlcolor=Chartreuse3]{hyperref}
\usepackage{acronym}
\usepackage{needspace}
\usepackage{float}
\usepackage[usenames,dvipsnames]{xcolor}
\usepackage{xspace}
\usepackage{tikz}
\usetikzlibrary{decorations.pathmorphing}
\usetikzlibrary{calc}
\usepackage{siunitx}
\usepackage{makecell}
\usepackage{multirow}
\usepackage{bm}
\usepackage{orcidlink}
\usepackage[caption=false]{subfig}
\usepackage{ulem}
\usepackage{booktabs}
\usepackage{physics}
\usepackage{tensor}
\usepackage{braket}
\usepackage{longtable}
\usepackage{array}
\usepackage{needspace}

\tikzset{
  worldline/.style={black, thick, line cap=rect},
  A/.style={black, thick, dash pattern=on 3pt off 3pt, line cap=rect},
  phi/.style={black, thick, line cap=rect},
  sigma/.style={black, thick, double, double distance=1.2pt, line cap=butt}
}

\allowdisplaybreaks

\begin{document}

\title{N-body next-to-leading order gravitational spin-orbit\\ 
interaction via effective field theory}
\author{Leonardo Wimmer\orcidlink{0009-0000-5503-8178}}
\email[Email: ]{wimmer@icrr.u-tokyo.ac.jp}
\affiliation{Institute for Cosmic Ray Research, The University of Tokyo, 5-1-5 Kashiwanoha, Kashiwa, Chiba 277-8582, Japan}
\author{Hideyuki Tagoshi\orcidlink{0000-0001-8530-9178}}
\affiliation{Institute for Cosmic Ray Research, The University of Tokyo, 5-1-5 Kashiwanoha, Kashiwa, Chiba 277-8582, Japan}

\begin{abstract}
Using the post-Newtonian effective field theory (PN-EFT) formalism for spinning gravitating bodies, we derive the next-to-leading-order (NLO) spin-orbit potential and Hamiltonian for a system of $N$ spinning bodies in general relativity. This extends the EFT treatment of the binary case to arbitrary $N$. We present two derivations: one in the generalized canonical gauge, and one based on the covariant spin supplementary condition (SSC), followed by a noncanonical transformation to canonical variables. In both approaches, the only new contributions beyond the binary case are three-body interaction diagrams. The canonical Hamiltonians obtained from the two EFT routes agree with the known ADM $N$-body Hamiltonian of Hartung and Steinhoff up to a canonical transformation.
\end{abstract}

\date{\today}

\maketitle

\section{Introduction}

Strong gravitational-wave (GW) signals are produced during the coalescence of spinning compact objects such as black holes (BHs) and neutron stars.
Since the first direct detection of GWs by the LIGO and Virgo Collaborations~\citep{gw150914}, the number of reported GW events has grown into the hundreds~\citep{gwtc2023,gwtc2025}.
KAGRA has since joined the global detector network, further expanding the observational reach of GW astronomy~\citep{ligo2015,virgo2015,kagra2021,lvk}.

To date, all confidently detected GW signals are consistent with compact binary coalescences (CBCs). Their detection and interpretation rely on accurate GW waveforms. For compact binaries, the inspiral is the longest stage of the signal, during which the bodies have non-relativistic velocities and remain well separated. Post-Newtonian (PN) theory describes this regime through a perturbative expansion in the slow-motion and weak-field limit, and provides the dynamics used to construct inspiral waveforms~\citep{PNblanchet2024}.

Several formalisms have been developed to derive compact-object dynamics in PN theory~\citep{PNblanchet2024}, including the ADM Hamiltonian formalism and the effective field theory (EFT) formalism.
The ADM approach to PN dynamics has a long history~\citep{kimura1961fixation,hiida1972gauge,ohta1974coord,ohta1974higher}, with later developments in~\citep{damour2008,steinhoff2008adm,hartung2011,js2012,js2013,js2015}.
The EFT approach to PN dynamics was introduced by Goldberger and Rothstein~\citep{goldberger2006,goldberger2006towers}.
For reviews, see~\citep{porto2016,levi2020}.

The EFT approach provides a systematic framework for computing both the conservative dynamics and radiation emission of a system of compact objects in the classical limit of gravity. The construction is based on the hierarchy of scales in the inspiral problem, which allows one to systematically build effective actions at the relevant scales order by order in the PN scheme. Although originally introduced as nonrelativistic general relativity (NRGR), we refer to this framework as PN-EFT~\citep{galleyhu2009,bernard2025}.

This formulation brings standard field-theory tools into classical gravity, including Feynman diagrams, matching, dimensional regularization, renormalization, and renormalization-group methods~\citep{goldberger2006,goldbergerross2010,galley2016,porto2016,levi2020}.
It also connects PN calculations with modern multi-loop and amplitude-based methods for compact-binary dynamics~\citep{foffa2017,buonannoEFT2022}.

Much effort has been devoted to high-order binary conservative dynamics in the nonspinning point-mass sector within PN-EFT.
In this sector, complete results have been obtained through 4PN order~\citep{foffa2019a,foffa2019b,blumlein2020a}, building on earlier partial results and analyses of tail effects and ambiguity issues~\citep{foffa2013,foffa2013tail,galley2016,foffa2017,porto2017lamb,porto2017}.
At 5PN order, potential-region contributions and local-in-time tail contributions have been computed~\citep{blumlein2021five2,blumlein2022five1}, with independent static-sector calculations in~\citep{foffa2019static,blumlein2020static}.
Far-zone hereditary effects have also been analyzed in~\citep{foffa2020hereditary,foffa2021near,almeida2021tail,almeida2023}, and scattering observables have recently been used to isolate further 5PN conservative contributions~\citep{porto2026five}.
Further progress at and beyond 5PN includes factorization methods for large classes of 5PN diagrams~\citep{foffa2021efficient}, partial 6PN results~\citep{blumlein2020six,blumlein2021sixG4}, and high-loop computations of the static sector at 6PN and 7PN order~\citep{brunello2025six,brunello2026seven}.

The PN-EFT approach was first extended to include spin and higher-order multipoles in~\citep{porto2006,porto2006eih}. Since then, the NLO spin-orbit and spin-spin interactions were derived in~\citep{porto2006eih,porto2008a,porto2008b,levi2010ss,levi2010so}, and NNLO results were obtained in~\citep{levi2015spinning,levi2012,levi2014,levi2015finite,levi2016so,levi2016ss}. The NNNLO spin-orbit and spin-spin sectors have also been studied in~\citep{kim2023a,kim2023b,mandal2023a,mandal2023b}, while higher powers in the spins, including cubic- and quartic-in-spin interactions, were derived in~\citep{levi2021cubic,levi2021quartic}. These developments have pushed the EFT treatment of spinning compact binaries to the 5PN frontier~\citep{levi2023fifthPN}. 
This also includes finite-size effects, which were treated in~\citep{porto2008a,levi2015finite,levi2015spinning}.

Looking ahead, future upgrades of LIGO, Virgo, and KAGRA, together with new detectors such as the space-based mission LISA~\cite{LISA2024} and third-generation ground-based observatories such as the Einstein Telescope~\cite{ET2020} and Cosmic Explorer~\cite{CE2021}, will improve GW sensitivity and extend the accessible frequency range.
This will make it possible to observe longer portions of the inspiral and characterize compact binaries with higher precision.
This is especially relevant for systems affected by additional companions.
In particular, LISA will observe compact binaries in the low-frequency band, roughly $10^{-4}\,\mathrm{Hz}$--$1\,\mathrm{Hz}$, where slow perturbations from a third body can accumulate over long observation times.

Many studies have explored how a third body can leave observable signatures in GW signals, especially for sources in the LISA band. 
A common scenario is a hierarchical system in which a compact binary orbits a massive, often supermassive, BH in a galactic nucleus or an AGN environment~\cite{cardoso,dorazio,deme,hoang,knee,liu1,sberna,wang,wong,xuan,yin,santos}. 
Such systems are astrophysically motivated: massive galaxies commonly host central supermassive black holes (SMBHs), and galactic nuclei can provide dense stellar environments where compact binaries interact with a central massive object~\cite{kormendy2013,graham2009,neumayer2020}.

The resulting GW signatures can take several forms. 
Secular perturbations can drive eccentricity oscillations through the Kozai--Lidov mechanism~\cite{chan,deme,gupta,hoang,knee,wang}, or excite eccentricity through apsidal-precession resonances~\cite{liu1,liu2}. 
The motion of the inner binary around the tertiary can also produce acceleration-induced dephasing, Doppler modulation, and Shapiro time delay~\cite{meiron,wong,xuan,sberna}, while lensing can lead to repeated amplification, wave-optics effects, and polarization-dependent propagation~\cite{dorazio,pijnenburg,oancea,santos}.
Related studies have also considered hierarchical triples without a central SMBH, including stellar triples, compact-object triples, and double-white-dwarf formation channels~\cite{meiron,robson,raja,liu2}. 
In strong-field configurations, a stellar-mass compact binary orbiting a massive BH can also modulate the signal and resonantly excite BH quasinormal modes (QNMs)~\cite{cardoso,santos,yin}.
Altogether, these studies suggest that third bodies can affect GW signals through phase, amplitude, polarization, and timing modulations, burst-like signatures, and resonant QNM excitation.

This motivates the development of waveform models for compact-object systems beyond isolated binaries. 
Since PN-EFT provides the conservative dynamics used in inspiral waveform modeling, it is natural to extend this framework to systems with more than two bodies. 
In the diagrammatic approach, this amounts to including diagrams with more than two worldlines.

Although this extension is conceptually straightforward, the literature beyond binaries remains limited.
Within PN-EFT, the general $N$-body problem has been treated for nonspinning point masses through 2PN order using perturbative field theory~\cite{chu}.
Beyond PN-EFT, amplitude-based post-Minkowskian (PM) methods have been used to study many-body dynamics, including conservative Hamiltonians for $N$ nonspinning bodies~\cite{jones} and the three-body problem at 2PM order in generic and hierarchical configurations~\cite{loebbert,solon}.
Spin-related effects beyond binaries have been studied in specialized hierarchical systems~\cite{fang,kuntz1,kuntz2}.
In particular, the spin-orbit dynamics of $N$ spinning compact bodies has not been derived in PN-EFT.

%Goal
In this paper, we derive the $N$-body NLO spin-orbit potential and Hamiltonian in the PN-EFT formalism, extending the binary results of~\cite{levi2010so,levi2015spinning} to arbitrary $N$. We perform the derivation in two ways: first in the generalized canonical gauge of~\cite{levi2015spinning}, and second using the covariant SSC approach of~\cite{levi2010so}, followed by a noncanonical transformation to canonical variables. As a consistency check, we show that the resulting canonical Hamiltonians agree with the ADM $N$-body Hamiltonian of Hartung and Steinhoff~\cite{hartung2011} after a canonical transformation.

%Outline
The paper is organized as follows. In Sec.~\ref{sec:II}, we review the PN-EFT formalism for spinning gravitating bodies and introduce the conventions used throughout the paper. In Sec.~\ref{sec:III}, we derive the $N$-body NLO spin-orbit potential from the relevant Feynman diagrams in the generalized canonical gauge. In Sec.~\ref{sec:IV}, we obtain the corresponding Hamiltonian and compare it with the ADM result of Hartung and Steinhoff. In Sec.~\ref{sec:V}, we repeat the calculation using the covariant SSC, deriving the required spin vertices, diagrams, potential, and Hamiltonian, before mapping the result to canonical variables and comparing it again with the ADM result. We conclude in Sec.~\ref{sec:VI}.

%Notation
Throughout this paper, we set $c=1$, use the mostly-minus signature, and denote $g\equiv\det g_{\mu\nu}$. Greek indices $\mu,\nu,\ldots$ denote 4-dimensional indices in the global coordinate frame running from $0$ to $3$. Lowercase Latin indices from the middle of the alphabet $i,j,k,\ldots$ denote spatial coordinate indices running from $1$ to $3$. Lowercase Latin letters from the beginning of the alphabet $a,b,c,\ldots$ denote body labels. Capital Latin indices $A,B,\ldots$ refer to the body-fixed Lorentz frame, whereas sans-serif Latin indices $\mathsf{a},\mathsf{b},\ldots$ refer to the Lorentz frame defined by the background tetrad. Both run from $0$ to $3$. We use $\int_k\equiv \int d^4k/(2\pi)^4$ and $\int_{\bm{k}}\equiv \int d^3\bm{k}/(2\pi)^3$. Boldface characters denote 3-vectors. Repeated tensor indices are summed over, whereas sums over body labels are written explicitly. We write scalar triple products as $\bm{a}\cdot\bm{b}\times\bm{c}\equiv \bm{a}\cdot(\bm{b}\times\bm{c})$.

\section{PN-EFT Formalism for Spinning Gravitating Bodies}
\label{sec:II}

In this section we summarize the PN-EFT formalism for gravitating spinning bodies \cite{levi2015spinning}, which we use to derive the $N$-body NLO spin-orbit interaction.

\subsection{Hierarchy of scales}

The PN-EFT framework is based on the separation of scales in the inspiral regime.
Consider a bound system of $N$ spinning compact bodies in the inspiral regime, with typical orbital velocity $v\ll 1$, typical mass $m$, and large orbital separations. 
The relevant scales are the internal scale of each compact body, $r_s\sim Gm$, the orbital separation scale $r$, and the radiation wavelength scale $\lambda$. 
For a gravitationally bound system, the virial theorem gives
\begin{equation}
    v^2\sim \frac{Gm}{r},
\end{equation}
while the orbital frequency satisfies $\omega_{\rm orb}\sim v/r$. 
Therefore, $r_s\sim r v^2$ and $\lambda\sim \omega_{\rm orb}^{-1}\sim r/v$, so that
\begin{equation}
    r_s \ll r \ll \lambda .
\end{equation}
This hierarchy motivates the EFT description. The dynamics of the system is obtained by integrating out one scale at a time, from the internal scale of each compact body to the orbital scale, and then to the radiation scale. At each step, one constructs an effective theory containing only the degrees of freedom relevant at the next scale, in a ``tower-like'' fashion. Since $r_s/r\sim v^2$ and $r/\lambda\sim v$, the dynamics is naturally organized as an expansion in the PN parameter $v$.

\subsection{The action}

After integrating out the internal scale of each compact object, one obtains a point-particle effective action valid at the orbital scale~\cite{levi2020,levi2015spinning}.
For a system of $N$ spinning compact bodies, this action is
\begin{align}
    S_{\text{GR}}
    &=
    S_{\rm bulk}[g_{\mu\nu}]
    +
    \sum_{a=1}^{N}
    S_{\text{pp}}^{(a)}
    \big[g_{\mu\nu},x_a^\mu,e_{aA}^{\mu}\big],
\end{align}
where $S_{\rm bulk}$ is the pure gravitational action and $S_{\rm pp}^{(a)}$ is the point-particle worldline action of the $a$-th spinning body. The point-particle degrees of freedom are the worldline $x_a^\mu(\lambda_a)$ and the body-fixed tetrad $e_{aA}^{\mu}(\lambda_a)$, both parametrized by the worldline parameter $\lambda_a$.

The pure gravitational action is given by the Einstein-Hilbert term together with a gauge-fixing term. 
We work in harmonic gauge ($\Gamma^\mu=0$), so that
\begin{align}
S_{\rm bulk}  = -\frac{1}{16\pi G}\int d^4x\,\sqrt{-g}\,\left(R-\frac{1}{2}g_{\mu\nu}\Gamma^\mu \Gamma^\nu\right),
\end{align}
where $ \Gamma^\mu \equiv g^{\rho\sigma}\Gamma^\mu_{\rho\sigma}$. Here $G$ is Newton's constant, $R$ is the Ricci scalar, and $g$ is the determinant of $g_{\mu\nu}$.

The point-particle worldline action for a spinning body is separated into a mass term and a spin-dependent term,
\begin{equation}
    S_{\text{pp}}=S_{\text{pp}(m)}\big[g_{\mu\nu},x^\mu\big]+S_{\text{pp}(\bm{S})}\big[g_{\mu\nu},x^\mu,e_A^\mu\big].
\end{equation}
The mass term is
\begin{equation}
    S_{\text{pp}(m)}
    =-m\!\int d\lambda \,\sqrt{u^2}
    =-m\!\int dt\, \big(g_{\mu\nu}v^\mu v^\nu\big)^{1/2},
\end{equation}
where $m$ is the mass of the body, $u^\mu \equiv dx^\mu/d\lambda$, and $u^2\equiv g_{\mu\nu}u^\mu u^\nu$.
The second equality follows from choosing the coordinate time $t=x^0$ as the worldline parameter and defining $v^\mu \equiv dx^\mu/dt$.
The spin term is discussed in the next subsection.

\subsection{Spin}

Tetrad are sets of four orthonormal vector fields defining a local Lorentz frame at each point of spacetime.
We use the tetrad conventions of~\cite{levi2010so}.
A spinning point particle is described by two tetrads on its worldline: a body-fixed tetrad $e_A^\mu$ and a background tetrad $e_{\mathsf a}^\mu$.
By definition, they satisfy
\begin{align}
    g_{\mu\nu} e_A^\mu e_B^\nu
    =
    \eta_{AB},
    \qquad
    g_{\mu\nu} e_{\mathsf{a}}^\mu e_{\mathsf{b}}^\nu
    =
    \eta_{\mathsf{a}\mathsf{b}} .
\end{align}
The body-fixed tetrad carries the rotational degrees of freedom of the spinning body, while the background tetrad provides the local Lorentz frame with respect to which this rotation is described.
The two tetrads are therefore related by a Lorentz transformation,
\begin{equation}
    e^A_\mu
    =
    \Lambda^A_{\, \mathsf{a}}\, e^{\mathsf{a}}_\mu ,
\end{equation}
where $\eta_{AB}\Lambda^A_{\,\mathsf{a}}\Lambda^B_{\,\mathsf{b}}=\eta_{\mathsf{a}\mathsf{b}}$ and $\eta^{\mathsf{a}\mathsf{b}}\Lambda^A_{\,\mathsf{a}}\Lambda^B_{\,\mathsf{b}}=\eta^{AB}$.
The Lorentz matrix $\Lambda^A_{\, \mathsf{a}}$ describes the orientation of the body-fixed frame with respect to the local frame defined by the background tetrad.

The angular velocity associated with the body-fixed tetrad is
\begin{equation}
    \Omega^{\mu\nu}
    =
    e_A^\mu
    \frac{D e^{A\nu}}{d\lambda},
\end{equation}
where $D/d\lambda$ is the covariant derivative along the worldline.
The orthonormality of the tetrad implies that $\Omega^{\mu\nu}$ is antisymmetric.
This definition generalizes the flat-spacetime expression~\cite{hanson1974}
\begin{equation}
    \Omega_{\rm flat}^{\mathsf{ab}}
    =
    \Lambda^\mathsf{a}_{\,A}
    \frac{d\Lambda^{A\mathsf{b}}}{d\lambda}.
\end{equation}
Projecting the angular velocity onto the local Lorentz frame gives
\begin{equation}
    \Omega^{\mathsf{ab}}
    =
    \Omega^{\mu\nu}e^\mathsf{a}_\mu e^\mathsf{b}_\nu
    =
    \Omega^{\mathsf{ab}}_{\rm flat}
    +u^\mu \omega_\mu^{\mathsf{ab}},
\end{equation}
where $\omega_\mu^{\mathsf{ab}}\equiv e^{\mathsf{b}}_{\nu}\nabla_\mu e^{\mathsf{a}\nu}$ is the spin connection.

The background tetrad is defined only up to a local Lorentz transformation.
We fix this freedom by choosing the symmetric tetrad whose local Lorentz frame coincides asymptotically with the coordinate frame, so that $e^\mathsf{a}_\mu\to\delta^\mathsf{a}_\mu$ when $h_{\mu\nu}\to0$.
For $\eta_{\mu\nu}+h_{\mu\nu}=\eta_{\mathsf{ab}}e^\mathsf{a}_\mu e^\mathsf{b}_\nu$, this gives
\begin{equation}
    e^\mathsf{a}_\mu
    =
    \delta^\mathsf{a}_\mu
    +\frac{1}{2}h^\mathsf{a}_{\,\mu}
    -\frac{1}{8}h^\mathsf{a}_{\,\rho}h^\rho_{\,\mu}
    +\cdots .
\end{equation}

The conjugate momentum and spin tensor are defined from the minimal-coupling part of the point-particle Lagrangian by
\begin{equation}
    p_\mu
    =
    -\frac{\partial L_{\rm pp}}{\partial u^\mu},
    \qquad
    S_{\mu\nu}
    =
    -2
    \frac{\partial L_{\rm pp}}{\partial \Omega^{\mu\nu}},
\end{equation}
where $L_{\rm pp}=L_{\rm pp}(u^\mu,\Omega^{\mu\nu})$.
The covariant spin tensor contains more components than the physical spin.
Since $S_{\mu\nu}$ is antisymmetric in four dimensions, it has 6 independent components.
These components correspond to the Lorentz rotational variables of the body-fixed frame.
For a massive spinning particle, however, the physical spin is described only by the 3 rotational degrees of freedom of the spatial triad.
The remaining 3 components are therefore redundant.

There are two main treatments of the redundant components of the spin tensor in spin-sector calculations in the PN-EFT literature.
In the formulation of~\cite{levi2015spinning}, one works in the generalized canonical gauge, where the spin gauge is fixed so that the spin variables are canonical.

In the earlier formulation of~\cite{levi2010so}, the EFT calculation is carried out with redundant spin variables, and the covariant spin supplementary condition (SSC) is used to eliminate the extra spin degrees of freedom.
To obtain a Hamiltonian in canonical variables, one then performs a noncanonical transformation from the covariant-SSC variables to the Pryce--Newton--Wigner SSC variables, which are known to be canonical.

The formulation in the generalized canonical gauge is more efficient, since canonical variables are used from the beginning.
Nevertheless, both treatments should give equivalent results for the spin-orbit dynamics.
In this paper, we use the generalized canonical gauge in Secs.~\ref{sec:III} and~\ref{sec:IV}.
In Sec.~\ref{sec:V}, we instead use redundant spin variables and impose the SSCs explicitly.
We describe these two approaches in more detail below.

\subsubsection{Generalized canonical gauge}

\newpage
In the generalized canonical gauge, the spin gauge is fixed from the start, so the spin couplings are written only in terms of the 3 independent components of $S^{ij}$.
Assuming minimal coupling, and after fixing the spin gauge of the rotational variables, the spin-dependent part of the worldline action in this gauge is given by~\cite{levi2015spinning}
\begin{widetext}
\begin{align}
    S_{\text{pp}(\bm{S})}
    &=
    - \int d\lambda \bigg[\,
    \frac{1}{2} S_{\mu\nu}\Omega^{\mu\nu}
    +\frac{S^{\mu\nu}p_\nu}{p^2}
    \frac{Dp_\mu}{d\lambda}
    \bigg] \notag\\
    &=
    -\int dt\, \bigg[
    \frac{1}{2}S^{ij}\Omega_{\mathrm{flat}}^{ij}
    -L_{\text{kin}}
    +\frac{1}{2}\omega_\mu^{ij}u^\mu
    \left(
    S^{ij}
    -\left(1+\frac{3}{4}v^2\right)S^{ik}v^k v^j
    \right)
    +\omega_\mu^{0i}u^\mu
    \left(1+\frac{v^2}{2}\right)S^{ij}v^j
    \bigg].
\end{align}
\end{widetext}
In the second line, we have chosen the coordinate time $t$ as the worldline parameter, with $v^i\equiv dx^i/dt$ and $a^i\equiv dv^i/dt$. Only the terms needed at the relevant orders are shown. The term $L_{\rm kin}$ is
\begin{equation}
    L_{\rm kin}
    =\frac{1}{2}\left(1+\frac{3}{4}v^2\right)
    S^{ij}v^i a^j .
    \label{lkin}
\end{equation}
Although this term does not contain NRG fields, it is linear in spin and contributes to the spin-orbit interaction at both LO and NLO.

\subsubsection{Spin supplementary conditions}

Alternatively, one can keep both $S^{ij}$ and the 3 redundant components $S^{0i}$ during the calculation, and remove the latter by imposing an SSC.
The spin action takes the form~\cite{levi2010so}
\begin{align}
    S_{\text{pp}(\bm{S})}&=-\int d\lambda \frac{1}{2}S_{\mu\nu}\Omega^{\mu\nu}
    \notag\\
    &=-\int d\lambda \,\bigg(
    \frac{1}{2}S_{\mathsf{ab}} \Omega^{\mathsf{ab}}_{\text{flat}}
    +\frac{1}{2}S_{\mathsf{ab}}\omega_\mu^{\mathsf{ab}}u^\mu
    \bigg).
\end{align}

Among the many different SSCs, the covariant SSC is given by $S^{\mu\nu}p_\nu=0$~\cite{tulczyjew1959}.
At linear order in spin, preserving the covariant SSC along the worldline,
$D(S^{\mu\nu}p_\nu)/d\lambda=0$, and using the Mathisson--Papapetrou
equations gives $p^\mu=m u^\mu/\sqrt{u^2}+\mathcal{O}(S^2)$~\cite{levi2010so}.
Thus, this condition can equivalently be written, to the relevant order for the spin-orbit sector, as $S^{\mu\nu}u_\nu=0$, or 
\begin{equation}
    S^{\mathsf{ab}} e_{\mathsf{b} \nu}u^\nu=0.
    \label{covssc}
\end{equation}
The Pryce--Newton--Wigner SSC~\cite{pryce1948, newton1949} is the flat-spacetime choice associated with canonical spin variables. It was generalized to curved spacetime in~\cite{barausse2009, steinhoff2008adm}, and shown to give canonical variables at linear order in spin~\cite{barausse2009}. It can be written as~\cite{levi2010so}
\begin{equation}
    S^{\mu\nu}\left(p_\nu+me^0_\nu\right)=0.
\end{equation}

\subsubsection{Power counting}

The PN counting of spin follows from the relation $S\sim m v_{\rm rot} r_s$, where $v_{\rm rot}$ is the rotational velocity of the compact body and $r_s\sim Gm$ its size. For maximally rotating compact objects, $v_{\rm rot}\lesssim 1$. In the inspiral regime, the virial theorem gives $Gm/r\sim v^2$, and therefore $r_s\sim rv^2$. Hence
\begin{equation}
    \frac{S}{mr}\sim v^2.
\end{equation}
In this sense, each spin counts as order $v^2$ in the PN expansion. 
The $\mathrm{N}^n\mathrm{LO}$ spin-orbit sector then contributes at $(3/2+n)$PN order. In particular, the LO spin-orbit sector enters at $1.5$PN order, while the NLO spin-orbit sector considered here enters at $2.5$PN order.

\subsection{The effective action}

In this section we explain how Feynman diagrams arise from the radiation-scale effective action in the PN-EFT formalism, and how the Feynman rules are computed.

\subsubsection{Feynman diagrams}

In the PN regime, the gravitational field is weak, so we expand the metric around a flat background. We write
\begin{equation}
    g_{\mu\nu}=\eta_{\mu\nu}+H_{\mu\nu}+\bar{h}_{\mu\nu},
\end{equation}
where $H_{\mu\nu}$ denotes potential gravitons and $\bar{h}_{\mu\nu}$ radiation gravitons. They scale as $(k_0,\bm{k})\sim (v/r,1/r)$ and $(k_0,\bm{k})\sim (v/r,v/r)$, respectively. Potential gravitons mediate the conservative orbital interaction, while radiation gravitons describe radiation emission~\cite{porto2016}.

The conservative dynamics is obtained by integrating out the potential modes $H_{\mu\nu}$ and setting the radiation modes to zero. This gives the effective action
\begin{equation}
    e^{iS_{\text{eff}}[J]}
    =
    \int D H_{\mu\nu}\,
    e^{iS_{\text{GR}}[g_{\mu\nu}=\eta_{\mu\nu}+H_{\mu\nu},J]} .
    \label{integrateout}
\end{equation}
Here $J=\{m_a,\bm{x}_a(t),\bm{S}_a(t)\}$ denotes the set of worldline source variables after choosing the coordinate time $t$ as the worldline parameter.
We use the Kol--Smolkin (KS) parametrization~\cite{kol2008} to write $g_{\mu\nu}=\eta_{\mu\nu}+H_{\mu\nu}$ in terms of non-relativistic gravitational (NRG) fields $\phi$, $A_i$, and $\sigma_{ij}$. 
Explicitly,
\begin{equation}
g_{\mu\nu}
= e^{2\phi}
\begin{pmatrix}
1 & -A_{j}\\[3pt]
-A_{i} & \, -\gamma_{ij}e^{-4\phi}+A_{i}A_{j}
\end{pmatrix},
\label{k-k}
\end{equation}
where $\gamma_{ij}\equiv \delta_{ij}+\sigma_{ij}$.
Then, we separate $S_{\text{GR}}$ into a kinetic term and a field-dependent term,
\begin{equation}
    S_{\text{GR}}[g_{\mu\nu}, J]=S_{\text{kin}}[J]+S_{\text{fields}}[g_{\mu\nu}, J].
\end{equation}
The functional integral is defined as
\begin{equation}
    Z[J] = \int D\phi\,D A_i\,D\sigma_{ij}\,e^{iS_{\text{fields}}[\phi, A_i,\sigma_{ij},J]}.
\end{equation}
In the classical limit,
\begin{equation}
    W[J]=-i\log Z[J] 
\end{equation}
is the sum of tree-level connected diagrams~\cite{porto2016}, computed using the Feynman rules in Table~\ref{tab:Feynman}.  
Here, ``connected'' means the diagrams cannot be decomposed into independent pieces once the source worldlines are removed, and ``tree level'' means no closed graviton loops, which would correspond to quantum corrections.

Finally, Eq.~\eqref{integrateout} yields
\begin{equation}
    S_{\text{eff}}[J]=S_{\text{kin}}[J]+W[J].
\end{equation}
Multiplying the Feynman rules while taking Wick contractions of the fields gives the contribution of each diagram to $W$. Symmetry factors must be taken into account. Following~\cite{levi2010so,levi2015spinning}, we present these results as Lagrangian contributions by stripping off the overall time integral, 
\begin{equation}
    W^{(\mathrm{diag})}=\int dt\, L^{(\mathrm{diag})}.
\end{equation}
The effective Lagrangian describing the conservative dynamics of the inspiral is then
\begin{equation}
    L_{\mathrm{eff}}=K_{\mathrm{eff}}-V_{\rm eff},
\end{equation}
where $K_{\text{eff}}$ is the kinetic term, $S_{\rm kin} = \int dt\,K_{\rm eff}$, and $V_{\rm eff}$ is the effective potential,
\begin{equation}
    V_{\rm eff}=-\sum_{\mathrm{diag}}L^{(\mathrm{diag})}.
\end{equation}
Thus, the interaction potential is obtained by evaluating all diagrams that contribute at the desired PN order.

\subsubsection{Feynman rules}

The Feynman rules used to compute $W$ are obtained by substituting the KS parametrization into $S_{\rm GR}$ and expanding in the NRG fields. We first consider the bulk term. After two integrations by parts, the part quadratic in the fields is
\begin{align}
    S_{\rm bulk}^{(\rm quad)}
    &=
    \frac{1}{32\pi G}
    \int \!d^4x\,
    \bigg[
    \!-4(\partial_i\phi)^2+4(\partial_0 \phi)^2
    +(\partial_i A_j)^2
    \notag\\
    &\qquad\qquad\;\;
    -(\partial_0A_i)^2
    +\frac{1}{4}(\partial_i\sigma_{jj})^2
    -\frac{1}{2}(\partial_i \sigma_{jk})^2
    \notag\\
    &\qquad\qquad\;\;
    -\frac{1}{4}(\partial_0 \sigma_{ii})^2
    +\frac{1}{2}(\partial_0 \sigma_{ij})^2
    \,\bigg] .
\end{align}
There are no mixed quadratic terms between $\phi$, $A_i$ and $\sigma_{ij}$. Thus, in the KS parametrization, the NRG fields are diagonal at quadratic order, and the propagators can be read directly from this action. 

To obtain the conservative dynamics, we consider only potential gravitons, whose modes scale as $(k_0,\bm{k})\sim(v/r,1/r)$.
The propagator can then be expanded in powers of $k_0^2/\bm{k}^2\sim v^2$:
\begin{align}
    \int_k\frac{e^{-ik\cdot x}}{k^2}
    &=
    \int_{k_0}\int_{\bm{k}}
    \frac{e^{-i(k_0t-\bm{k}\cdot\bm{x})}}{k_0^2-\bm{k}^2}
    \notag\\
    &=
    -\int_{k_0}\int_{\bm{k}}
    \frac{e^{-i(k_0t-\bm{k}\cdot\bm{x})}}{\bm{k}^2}
    \left(1+\frac{k_0^2}{\bm{k}^2}+\cdots\right)
    \notag\\
    &=
    -\delta(t)\int_{\bm{k}}
    \frac{e^{i\bm{k}\cdot\bm{x}}}{\bm{k}^2}
    +\cdots.
\end{align}
At leading order, the propagators are instantaneous. The remaining terms are relativistic time corrections to the propagators. Following~\cite{levi2020}, we treat these corrections as separate quadratic vertices, which we call propagator-correction vertices. These vertices are denoted by a circled cross.

Cubic and higher-order terms in the bulk action generate self-interaction vertices for the NRG fields. The cubic terms needed for the NLO spin-orbit sector are
\begin{align}
    S_{\rm bulk}^{(\rm cubic)} 
    &\supset 
    \frac{1}{32\pi G}
    \!\int\! d^4x\,
    \Big[
    4\phi\big(\partial_i A_j\,(\partial_i A_j-\partial_j A_i)+(\partial_i A_i)^2 \big)
    \notag\\[2pt]
    &\qquad\qquad\quad\,\,
    +2\big(2\sigma_{ij}\partial_i \phi\, \partial_j \phi-\sigma_{ii}\partial_j \phi\, \partial_j \phi\big)
    \notag\\[2pt]
    &\qquad\qquad\quad\,\,
    -8 A_i \partial_i \phi\, \partial_0 \phi 
    \,\Big].
\end{align}
The relevant cubic vertices are $\phi AA$, $\sigma\phi\phi$, and $A\phi\phi$. The last one contains a time derivative and is denoted in the diagrams by a circled cross.

The worldline action produces two types of vertices: mass vertices and spin vertices. The mass vertices come from $S_{\text{pp}(m)}$, while the spin vertices come from $S_{\text{pp}(\bm{S})}$. Expanding these terms in the NRG fields gives
\begin{widetext}
\begin{align}
    S_{\text{pp}(m)}
    &=
    -m\!\int\! dt\,
    \Bigg[
    1-\frac{1}{2}v^2-\frac{1}{8}v^4+\cdots
    +\left(\phi+\frac{3}{2}\phi\, v^2-A_iv^i-\frac{1}{2}A_i v^i v^2-\frac{1}{2}\sigma_{ij}v^i v^j\right)
    \notag\\[2pt]
    &\qquad\qquad\qquad
    +\left(\frac{1}{2}\phi^2-\phi A_iv^i \right)
    +\cdots
    \Bigg],
    \\
    S_{\text{pp}(\bm{S})}
    &=
    \int \!dt\, S^{ij}
    \Bigg[
    2v^i \partial_j \phi
    + v^2 v^i \partial_j \phi
    -2 v^i a^j \phi
    +\frac{1}{2}\partial_i A_j
    +\frac{3}{4}v^i v^k(\partial_k A_j-\partial_j A_k)
    +v^i \partial_0 A_j
    \notag\\[2pt]
    &\qquad\qquad\qquad
    +\frac{1}{2}v^k\partial_i \sigma_{jk}
    +2\phi \partial_i A_j
    +\cdots
    \Bigg].
\end{align}
\end{widetext}
The first ellipsis above denotes higher-order kinetic terms, which are independent of the fields. The remaining ellipses denote field-dependent terms that do not contribute to the NLO spin-orbit sector.
In the diagrams, white circles represent mass vertices and black circles represent spin vertices.
Labels such as $v^n$ attached to vertices indicate the velocity scaling of the corresponding term in the Feynman rule.

All Feynman rules required to obtain the NLO spin-orbit interaction are collected in Table~\ref{tab:Feynman}, where we use the shorthand
\begin{equation}
    \mathcal{P}_{ijk\ell}
    =
    \frac{1}{2}
    \left(
    \delta_{ik}\delta_{j\ell}
    +\delta_{jk}\delta_{i\ell}
    -2\delta_{ij}\delta_{k\ell}
    \right).
\end{equation}

\begingroup
\setlength\LTleft{0pt}
\setlength\LTright{0pt}
\setlength\LTcapwidth{\textwidth}

\begin{longtable*}{l}
\caption{
Feynman rules in position space required for the NLO spin-orbit calculation. The worldline spin vertices are valid in the generalized canonical gauge. Besides these rules, only the symmetry factor is necessary to calculate any diagram.}
\label{tab:Feynman}\\
\hline\hline
\endfirsthead

\hline\hline
\endhead

\hline\hline
\endfoot

\hline\hline
\endlastfoot

%------------------ Propagators ------------------
\noalign{\vskip 2pt}
\multicolumn{1}{c}{Propagators} \\
\noalign{\vskip 1pt}
\hline
\noalign{\vskip 8pt}

\makebox[0.75\textwidth][c]{%
\parbox{0.75\textwidth}{\centering
$\displaystyle
\begin{aligned}
\vcenter{\hbox{
\begin{tikzpicture}[baseline=(current bounding box.center),x=0.8cm,y=0.8cm,line cap=round,line join=round]
  \draw[phi] (-1,0) -- (1,0);
\end{tikzpicture}}}
&\;=\;\,\,\,\,
\langle \phi(x_1)\phi(x_2)\rangle
&&\!\!\!\!\!=\; 4\pi G\,\int_{\bm{k}}
\frac{e^{i\bm{k}\cdot(\bm{x}_1-\bm{x}_2)}}{\bm{k}^2}\delta(t_1-t_2)
\\[4pt]
\vcenter{\hbox{
\begin{tikzpicture}[baseline=(current bounding box.center),x=0.8cm,y=0.8cm,line cap=round,line join=round]
  \draw[A] (-1,0) -- (1,0);
\end{tikzpicture}}}
&\;=\;\,
\langle A_i(x_1)A_j(x_2)\rangle
&&\!\!\!\!\!=\; -16\pi G\,\delta_{ij}\int_{\bm{k}}
\frac{e^{i\bm{k}\cdot(\bm{x}_1-\bm{x}_2)}}{\bm{k}^2}\delta(t_1-t_2)
\\[4pt]
\vcenter{\hbox{
\begin{tikzpicture}[baseline=(current bounding box.center),x=0.8cm,y=0.8cm,line cap=round,line join=round]
  \draw[sigma] (-1,0) -- (1,0);
\end{tikzpicture}}}
&\;=\;
\langle \sigma_{ij}(x_1)\sigma_{k\ell}(x_2)\rangle
&&\!\!\!\!\!=\; 32\pi G\,\mathcal{P}_{ijk\ell}\int_{\bm{k}}
\frac{e^{i\bm{k}\cdot(\bm{x}_1-\bm{x}_2)}}{\bm{k}^2}\delta(t_1-t_2)
\end{aligned}
$
}}\\[6pt]

%------------------ Propagator correction vertices ------------------
\noalign{\vskip 12pt}
\hline
\noalign{\vskip 2pt}
\multicolumn{1}{c}{Propagator-correction vertices} \\
\noalign{\vskip 1pt}
%\hline
\noalign{\vskip 8pt}

\multicolumn{1}{@{}c@{}}{%
\parbox{0.75\textwidth}{\centering
$\displaystyle
\begin{aligned}
\vcenter{\hbox{
\begin{tikzpicture}[baseline=(current bounding box.center),x=0.8cm,y=0.8cm,line cap=round,line join=round]
  \draw[phi] (-1,0) -- (1,0);
  \draw[black, line width=0.6pt, fill=white] (0,0) circle (3.5pt);
  \draw[black, line width=0.6pt] ($(0,0)+(-2.2pt,-2.2pt)$) -- ($(0,0)+(2.2pt,2.2pt)$);
  \draw[black, line width=0.6pt] ($(0,0)+(-2.2pt,2.2pt)$) -- ($(0,0)+(2.2pt,-2.2pt)$);
\end{tikzpicture}}}
&\;=\;
\frac{1}{8\pi G}\int d^4x\,\big(\partial_0\phi\big)^2
\\[4pt]
\vcenter{\hbox{
\begin{tikzpicture}[baseline=(current bounding box.center),x=0.8cm,y=0.8cm,line cap=round,line join=round]
  \draw[A] (-1,0) -- (1,0);
  \draw[black, line width=0.6pt, fill=white] (0,0) circle (3.5pt);
  \draw[black, line width=0.6pt] ($(0,0)+(-2.2pt,-2.2pt)$) -- ($(0,0)+(2.2pt,2.2pt)$);
  \draw[black, line width=0.6pt] ($(0,0)+(-2.2pt,2.2pt)$) -- ($(0,0)+(2.2pt,-2.2pt)$);
\end{tikzpicture}}}
&\;=\;
-\frac{1}{32\pi G}\int d^4x\,\big(\partial_0 A_i\big)^2
\\[4pt]
\vcenter{\hbox{
\begin{tikzpicture}[baseline=(current bounding box.center),x=0.8cm,y=0.8cm,line cap=round,line join=round]
  \draw[sigma] (-1,0) -- (1,0);
  \draw[black, line width=0.6pt, fill=white] (0,0) circle (3.5pt);
  \draw[black, line width=0.6pt] ($(0,0)+(-2.2pt,-2.2pt)$) -- ($(0,0)+(2.2pt,2.2pt)$);
  \draw[black, line width=0.6pt] ($(0,0)+(-2.2pt,2.2pt)$) -- ($(0,0)+(2.2pt,-2.2pt)$);
\end{tikzpicture}}}
&\;=\;
\frac{1}{128\pi G}\int d^4x\,
\Big[2(\partial_0\sigma_{ij})^2-(\partial_0\sigma_{ii})^2\Big]
\end{aligned}
$
}}\\[6pt]

%------------------ Worldline mass vertices ------------------
\noalign{\vskip 12pt}
\hline
\noalign{\vskip 2pt}
\multicolumn{1}{c}{Worldline mass vertices} \\
\noalign{\vskip 1pt}
\hline
\noalign{\vskip 8pt}

\multicolumn{1}{@{}c@{}}{%
\parbox{0.75\textwidth}{\centering
$\displaystyle
\begin{aligned}
&\vcenter{\hbox{
\begin{tikzpicture}[baseline=(current bounding box.center),x=0.8cm,y=0.8cm,line cap=round,line join=round]
  \draw[phi] (0,0) -- (1,0);
  \draw[worldline] (0,-0.8) -- (0,0.8);
  \draw[black, line width=0.6pt, fill=white] (0,0) circle (2.5pt);
\end{tikzpicture}}}
&&\!\!\!=\;
-m\!\int dt\,\phi
\left(1+\frac{3}{2}v^2\right)
\\[4pt]
&\vcenter{\hbox{
\begin{tikzpicture}[baseline=(current bounding box.center),x=0.8cm,y=0.8cm,line cap=round,line join=round]
  \draw[A] (0,0) -- (1,0);
  \draw[worldline] (0,-0.8) -- (0,0.8);
  \draw[black, line width=0.6pt, fill=white] (0,0) circle (2.5pt);
\end{tikzpicture}}}
&&\!\!\!=\;
m\!\int dt\,A_i\,v^i
\left(1+\frac{1}{2}v^2\right)
\\[4pt]
&\vcenter{\hbox{
\begin{tikzpicture}[baseline=(current bounding box.center),x=0.8cm,y=0.8cm,line cap=round,line join=round]
  \draw[sigma] (0,0) -- (1,0);
  \draw[worldline] (0,-0.8) -- (0,0.8);
  \draw[black, line width=0.6pt, fill=white] (0,0) circle (2.5pt);
\end{tikzpicture}}}
&&\!\!\!=\;
m\!\int dt\,\frac{1}{2}\sigma_{ij}\,v^iv^j
\\[4pt]
&\vcenter{\hbox{
\begin{tikzpicture}[baseline=(current bounding box.center),x=0.8cm,y=0.8cm,line cap=round,line join=round]
  \draw[worldline] (0,-0.8) -- (0,0.8);
  \draw[phi] (0,0) -- ({cos(30)},{sin(30)});
  \draw[phi] (0,0) -- ({cos(30)},{-sin(30)});
  \draw[black, line width=0.6pt, fill=white] (0,0) circle (2.5pt);
\end{tikzpicture}}}
&&\!\!\!=\;
-m\!\int dt\,\frac{1}{2}\phi^2
\\[4pt]
&\vcenter{\hbox{
\begin{tikzpicture}[baseline=(current bounding box.center),x=0.8cm,y=0.8cm,line cap=round,line join=round]
  \draw[worldline] (0,-0.8) -- (0,0.8);
  \draw[A] (0,0) -- ({cos(30)},{sin(30)});
  \draw[phi] (0,0) -- ({cos(30)},{-sin(30)});
  \draw[black, line width=0.6pt, fill=white] (0,0) circle (2.5pt);
\end{tikzpicture}}}
&&\!\!\!=\;
m\!\int dt\,\phi A_i\,v^i
\end{aligned}
$
}}\\[6pt]

%------------------ Worldline spin vertices ------------------
\noalign{\vskip 12pt}
\hline
\noalign{\vskip 2pt}
\multicolumn{1}{c}{Worldline spin vertices} \\
\noalign{\vskip 1pt}
\hline
\noalign{\vskip 8pt}

\multicolumn{1}{@{}c@{}}{%
\parbox{0.75\textwidth}{\centering
$\displaystyle
\begin{aligned}
&\vcenter{\hbox{
\begin{tikzpicture}[baseline=(current bounding box.center),x=0.8cm,y=0.8cm,line cap=round,line join=round]
  \draw[worldline] (0,-0.8) -- (0,0.8);
  \draw[phi] (0,0) -- (1,0);
  \draw[black, line width=0.6pt, fill=black] (0,0) circle (2.5pt);
\end{tikzpicture}}}
&&\!\!\!=
\int dt\,S^{ij}v^i\Big(
2\partial_j \phi+v^2 \partial_j\phi-2a^j \phi
\Big)
\\[4pt]
&\vcenter{\hbox{
\begin{tikzpicture}[baseline=(current bounding box.center),x=0.8cm,y=0.8cm,line cap=round,line join=round]
  \draw[worldline] (0,-0.8) -- (0,0.8);
  \draw[A] (0,0) -- (1,0);
  \draw[black, line width=0.6pt, fill=black] (0,0) circle (2.5pt);
\end{tikzpicture}}}
&&\!\!\!=
\int dt\,S^{ij}\bigg[
\frac{1}{2}\partial_i A_j+\frac{3}{4}v^iv^k (\partial_k A_j-\partial_j A_k)+v^i\partial_0 A_j
\bigg]
\\[4pt]
&\vcenter{\hbox{
\begin{tikzpicture}[baseline=(current bounding box.center),x=0.8cm,y=0.8cm,line cap=round,line join=round]
  \draw[worldline] (0,-0.8) -- (0,0.8);
  \draw[sigma] (0,0) -- (1,0);
  \draw[black, line width=0.6pt, fill=black] (0,0) circle (2.5pt);
\end{tikzpicture}}}
&&\!\!\!=
\int dt\,\frac{1}{2}S^{ij}v^k \partial_i \sigma_{jk}
\\[4pt]
&\vcenter{\hbox{
\begin{tikzpicture}[baseline=(current bounding box.center),x=0.8cm,y=0.8cm,line cap=round,line join=round]
  \draw[worldline] (0,-0.8) -- (0,0.8);
  \draw[A] (0,0) -- ({cos(30)},{sin(30)});
  \draw[phi] (0,0) -- ({cos(30)},{-sin(30)});
  \draw[black, line width=0.6pt, fill=black] (0,0) circle (2.5pt);
\end{tikzpicture}}}
&&\!\!\!=
\int dt\,2S^{ij}\phi\,\partial_iA_j
\end{aligned}
$
}}\\[6pt]

%------------------ Bulk vertices ------------------
\noalign{\vskip 12pt}
\hline
\noalign{\vskip 2pt}
\multicolumn{1}{c}{Bulk vertices} \\
\noalign{\vskip 1pt}
%\hline
\noalign{\vskip 8pt}

\multicolumn{1}{@{}c@{}}{%
\parbox{0.75\textwidth}{\centering
$\displaystyle
\begin{aligned}
\vcenter{\hbox{
\begin{tikzpicture}[baseline=(current bounding box.center),x=0.7cm,y=0.7cm,line cap=round,line join=round]
  \draw[phi] (-1.15,0) -- (0,0);
  \draw[A] (0,0) -- (0.8,0.8);
  \draw[A] (0,0) -- (0.8,-0.8);
\end{tikzpicture}}}
&=\;
\frac{1}{8\pi G}\int d^4x\,
\phi\Big[\partial_i A_j\big(\partial_i A_j-\partial_j A_i\big)+(\partial_iA_i)^2\Big]
\\[4pt]
\vcenter{\hbox{
\begin{tikzpicture}[baseline=(current bounding box.center),x=0.7cm,y=0.7cm,line cap=round,line join=round]
  \draw[sigma] (-1.15,0) -- (0.05,0);
  \draw[phi] (0,0) -- (0.8,0.8);
  \draw[phi] (0,0) -- (0.8,-0.8);
\end{tikzpicture}}}
&=\;
\frac{1}{16\pi G}\int d^4x\,
\Big(2\sigma_{ij}\partial_i\phi\partial_j\phi-\sigma_{ii}\partial_j\phi\partial_j\phi\Big)
\\[4pt]
\vcenter{\hbox{
\begin{tikzpicture}[baseline=(current bounding box.center),x=0.7cm,y=0.7cm,line cap=round,line join=round]
  \draw[A] (-1.15,0) -- (0,0);
  \draw[phi] (0,0) -- (0.8,0.8);
  \draw[phi] (0,0) -- (0.8,-0.8);
  \draw[black, line width=0.6pt, fill=white] (0,0) circle (3.5pt);
  \draw[black, line width=0.6pt] ($(0,0)+(-2.2pt,-2.2pt)$) -- ($(0,0)+(2.2pt,2.2pt)$);
  \draw[black, line width=0.6pt] ($(0,0)+(-2.2pt,2.2pt)$) -- ($(0,0)+(2.2pt,-2.2pt)$);
\end{tikzpicture}}}
&=\;
-\frac{1}{4\pi G}\int d^4x\,A_i\partial_i\phi\,\partial_0\phi
\end{aligned}
$
}}\\[6pt]

\noalign{\vskip 12pt}
\end{longtable*}
\endgroup

\section{NLO Spin-Orbit Potential}
\label{sec:III}

In this section, we derive the LO and NLO spin-orbit potentials for a system of $N$ spinning gravitating bodies.
At each order, we first list all generic worldline diagrams (diagrams with unassigned source worldlines) that contribute to the interaction.
We then attach the source worldlines in all possible ways. Diagrams with $2$ worldlines are called two-body diagrams, while diagrams with $3$ worldlines are called three-body diagrams.
Diagrams involving four or more bodies first enter the spin-orbit sector at higher orders and do not contribute here.
Therefore, the complete $N$-body spin-orbit interaction at LO and NLO is accounted for by two-body and three-body diagrams.

\subsection{LO diagrams}

First, we compute the LO spin-orbit diagrams. 
The LO spin-orbit sector receives contributions only at $\mathcal{O}(G)$. At $\mathcal{O}(G)$, the only contributing topology is the one-graviton exchange. The corresponding generic worldline diagrams are shown in Fig.~\ref{fig:LOSO_generic}. 
Since the diagrams contain only two source insertions, there are no possible three-body contributions at the LO spin-orbit order. The $N$-body result can be obtained directly from the known binary expressions in~\cite{levi2015spinning} by promoting the body labels to arbitrary particles and summing over all bodies.

\begin{figure}[H]
\centering
\begin{tabular}{@{}c@{\hspace{6mm}}c@{}}

\begin{tikzpicture}[x=1cm,y=1cm, line cap=round, scale=0.8, transform shape]
  % interaction
  \draw[phi] (2,0) -- (2,2);
  % worldline vertex
  \draw[black, line width=0.6pt, fill=white] (2,0) circle (2.5pt);
  % spin coupling
  \draw[black, line width=0.6pt, fill=black] (2,2) circle (2.5pt);
  % labels
  \node[above] at (2,2) {$v^3$};
  \node[below] at (2,0) {$v^0$};
\end{tikzpicture}

&

\begin{tikzpicture}[x=1cm,y=1cm, line cap=round, scale=0.8, transform shape]
  % interaction
  \draw[A] (2,0) -- (2,2);
  % worldline vertex
  \draw[black, line width=0.6pt, fill=white] (2,0) circle (2.5pt);
  % spin coupling
  \draw[black, line width=0.6pt, fill=black] (2,2) circle (2.5pt);
  % labels
  \node[above] at (2,2) {$v^2$};
  \node[below] at (2,0) {$v^1$};
\end{tikzpicture}

\end{tabular}
\caption{Generic worldline diagrams contributing to the LO spin-orbit interaction. The black and white circles denote spin and mass couplings, respectively.}
\label{fig:LOSO_generic}
\end{figure}
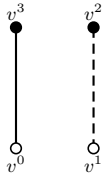

\begin{figure}[H]
\centering
\begin{tabular}{@{}c@{\hspace{6mm}}c@{}}

\begin{tikzpicture}[x=1cm,y=1cm, line cap=round, scale=0.8, transform shape]
  % interaction
  \draw[phi] (2,0) -- (2,2);
  % black lines
  \draw[worldline] (0.5,0) -- (3.5,0);
  \draw[worldline] (0.5,2) -- (3.5,2);
  % worldline vertex
  \draw[black, line width=0.6pt, fill=white] (2,0) circle (2.5pt);
  % spin coupling
  \draw[black, line width=0.6pt, fill=black] (2,2) circle (2.5pt);
  % labels
  \node[above] at (2,2) {$v^3$};
  \node[below] at (2,0) {$v^0$};
\end{tikzpicture}

&

\begin{tikzpicture}[x=1cm,y=1cm, line cap=round, scale=0.8, transform shape]
  % interaction
  \draw[A] (2,0) -- (2,2);
  % black lines
  \draw[worldline] (0.5,0) -- (3.5,0);
  \draw[worldline] (0.5,2) -- (3.5,2);
  % worldline vertex
  \draw[black, line width=0.6pt, fill=white] (2,0) circle (2.5pt);
  % spin coupling
  \draw[black, line width=0.6pt, fill=black] (2,2) circle (2.5pt);
  % labels
  \node[above] at (2,2) {$v^2$};
  \node[below] at (2,0) {$v^1$};
\end{tikzpicture}

\\[-1mm]
(a1) & (a2)

\end{tabular}
\caption{LO spin-orbit Feynman diagrams. The black and white circles denote spin and mass couplings, respectively.}
\label{fig:LOSO}
\end{figure}
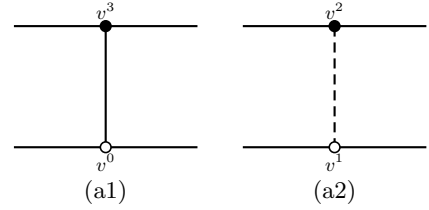

Attaching the source worldlines to the generic diagrams in Fig.~\ref{fig:LOSO_generic} gives the two diagrams shown in Fig.~\ref{fig:LOSO}. 
Then, using the Feynman rules in Table~\ref{tab:Feynman} yields
\begin{align}
    L^{(\text{a1})} &= \sum_a\sum_{b\neq a}\frac{2Gm_b}{r_{ab}^2}\bm{S}_a\!\cdot\bm{v}_a\times\bm{n}_{ab}, \\[2pt]
    L^{(\text{a2})} &= -\sum_a\sum_{b\neq a}\frac{2Gm_b}{r_{ab}^2}\bm{S}_a\!\cdot\bm{v}_b\times\bm{n}_{ab}.
\end{align}
We use the definitions
\begin{equation}
    \bm{r}_{ab}\equiv \bm{x}_a-\bm{x}_b,\qquad
    r_{ab}\equiv |\bm{r}_{ab}|,\qquad
    \bm{n}_{ab}\equiv \frac{\bm{r}_{ab}}{r_{ab}} .
\end{equation}
The spin vector is defined by $S^{ij}\equiv\varepsilon_{ijk}S^k$, or equivalently
\begin{equation}
    S^i=\frac{1}{2}\varepsilon_{ijk}S^{jk}.
\end{equation}

\subsection{NLO diagrams}

\begin{figure}[H]
\begin{center}

\begin{tabular}{@{}c@{\hspace{4mm}}c@{\hspace{4mm}}c@{\hspace{4mm}}c@{\hspace{4mm}}c@{\hspace{4mm}}c@{\hspace{4mm}}c@{}}

\begin{tikzpicture}[x=1cm,y=1cm, line cap=round, scale=0.8, transform shape]
  % interaction
  \draw[phi] (2,0) -- (2,2);
  % endpoint circles
  \draw[black, line width=0.6pt, fill=white] (2,0) circle (2.5pt);
  \draw[black, line width=0.6pt, fill=white] (2,2) circle (2.5pt);
  % spin coupling
  \draw[black, line width=0.6pt, fill=black] (2,2) circle (2.5pt);
  % labels
  \node[above] at (2,2) {$v^3$};
  \node[below] at (2,0) {$v^2$};
\end{tikzpicture}
&
\begin{tikzpicture}[x=1cm,y=1cm, line cap=round, scale=0.8, transform shape]
  % interaction
  \draw[phi] (2,0) -- (2,2);
  % endpoint circles
  \draw[black, line width=0.6pt, fill=white] (2,0) circle (2.5pt);
  \draw[black, line width=0.6pt, fill=white] (2,2) circle (2.5pt);
  % spin coupling
  \draw[black, line width=0.6pt, fill=black] (2,2) circle (2.5pt);
  % labels
  \node[above] at (2,2) {$v^5$};
  \node[below] at (2,0) {$v^0$};
\end{tikzpicture}
&
\begin{tikzpicture}[x=1cm,y=1cm, line cap=round, scale=0.8, transform shape]
  % interaction
  \draw[phi] (2,0) -- (2,2);
  % endpoint circles
  \draw[black, line width=0.6pt, fill=white] (2,0) circle (2.5pt);
  \draw[black, line width=0.6pt, fill=white] (2,2) circle (2.5pt);
  % spin coupling
  \draw[black, line width=0.6pt, fill=black] (2,2) circle (2.5pt);
  % propagator correction
  \draw[black, line width=0.6pt, fill=white] (2,1) circle (3.5pt);
  \draw[black, line width=0.6pt] ($(2,1)+(-2.2pt,-2.2pt)$) -- ($(2,1)+(2.2pt,2.2pt)$);
  \draw[black, line width=0.6pt] ($(2,1)+(-2.2pt, 2.2pt)$) -- ($(2,1)+(2.2pt,-2.2pt)$);
  % labels
  \node[above] at (2,2) {$v^3$};
  \node[below] at (2,0) {$v^0$};
\end{tikzpicture}
&
\begin{tikzpicture}[x=1cm,y=1cm, line cap=round, scale=0.8, transform shape]
  % interaction
  \draw[A] (2,0) -- (2,2);
  % endpoint circles
  \draw[black, line width=0.6pt, fill=white] (2,0) circle (2.5pt);
  \draw[black, line width=0.6pt, fill=white] (2,2) circle (2.5pt);
  % spin coupling
  \draw[black, line width=0.6pt, fill=black] (2,2) circle (2.5pt);
  % labels
  \node[above] at (2,2) {$v^2$};
  \node[below] at (2,0) {$v^3$};
\end{tikzpicture}
&
\begin{tikzpicture}[x=1cm,y=1cm, line cap=round, scale=0.8, transform shape]
  % interaction
  \draw[A] (2,0) -- (2,2);
  % endpoint circles
  \draw[black, line width=0.6pt, fill=white] (2,0) circle (2.5pt);
  \draw[black, line width=0.6pt, fill=white] (2,2) circle (2.5pt);
  % spin coupling
  \draw[black, line width=0.6pt, fill=black] (2,2) circle (2.5pt);
  % labels
  \node[above] at (2,2) {$v^4$};
  \node[below] at (2,0) {$v^1$};
\end{tikzpicture}
&
\begin{tikzpicture}[x=1cm,y=1cm, line cap=round, scale=0.8, transform shape]
  % interaction
  \draw[A] (2,0) -- (2,2);
  % endpoint circles
  \draw[black, line width=0.6pt, fill=white] (2,0) circle (2.5pt);
  \draw[black, line width=0.6pt, fill=white] (2,2) circle (2.5pt);
  % spin coupling
  \draw[black, line width=0.6pt, fill=black] (2,2) circle (2.5pt);
  % propagator correction
  \draw[black, line width=0.6pt, fill=white] (2,1) circle (3.5pt);
  \draw[black, line width=0.6pt] ($(2,1)+(-2.2pt,-2.2pt)$) -- ($(2,1)+(2.2pt,2.2pt)$);
  \draw[black, line width=0.6pt] ($(2,1)+(-2.2pt, 2.2pt)$) -- ($(2,1)+(2.2pt,-2.2pt)$);
  % labels
  \node[above] at (2,2) {$v^2$};
  \node[below] at (2,0) {$v^1$};
\end{tikzpicture}
&
\begin{tikzpicture}[x=1cm,y=1cm, line cap=round, scale=0.8, transform shape]
  % interaction
  \draw[sigma] (2,0) -- (2,2);
  % endpoint circles
  \draw[black, line width=0.6pt, fill=white] (2,0) circle (2.5pt);
  \draw[black, line width=0.6pt, fill=white] (2,2) circle (2.5pt);
  % spin coupling
  \draw[black, line width=0.6pt, fill=black] (2,2) circle (2.5pt);
  % labels
  \node[above] at (2,2) {$v^3$};
  \node[below] at (2,0) {$v^2$};
\end{tikzpicture}

\\
\end{tabular}

\vspace{0pt}

\begin{tabular}{@{}c@{\hspace{2mm}}c@{\hspace{2mm}}c@{\hspace{2mm}}c@{}}

\begin{tikzpicture}[x=1cm,y=1cm, line cap=round, scale=0.8, transform shape]
  % interaction
  \draw[A] (1,0) -- (2,2);
  \draw[phi] (3,0) -- (2,2);
  % endpoint circles
  \draw[black, line width=0.6pt, fill=white] (1,0) circle (2.5pt);
  \draw[black, line width=0.6pt, fill=white] (3,0) circle (2.5pt);
  \draw[black, line width=0.6pt, fill=white] (2,2) circle (2.5pt);
  % spin coupling
  \draw[black, line width=0.6pt, fill=black] (2,2) circle (2.5pt);
  % labels
  \node[above] at (2,2) {$v^2$};
  \node[below] at (1,0) {$v^1$};
  \node[below] at (3,0) {$v^0$};
\end{tikzpicture}
&
\begin{tikzpicture}[x=1cm,y=1cm, line cap=round, scale=0.8, transform shape]
  % interaction
  \draw[A] (1,0) -- (2,2);
  \draw[phi] (3,0) -- (2,2);
  % endpoint circles
  \draw[black, line width=0.6pt, fill=white] (1,0) circle (2.5pt);
  \draw[black, line width=0.6pt, fill=white] (3,0) circle (2.5pt);
  \draw[black, line width=0.6pt, fill=white] (2,2) circle (2.5pt);
  % spin coupling
  \draw[black, line width=0.6pt, fill=black] (1,0) circle (2.5pt);
  % labels
  \node[above] at (2,2) {$v^1$};
  \node[below] at (1,0) {$v^2$};
  \node[below] at (3,0) {$v^0$};
\end{tikzpicture}
&
\begin{tikzpicture}[x=1cm,y=1cm, line cap=round, scale=0.8, transform shape]
  % interaction
  \draw[phi] (1,0) -- (2,2);
  \draw[phi] (3,0) -- (2,2);
  % endpoint circles
  \draw[black, line width=0.6pt, fill=white] (1,0) circle (2.5pt);
  \draw[black, line width=0.6pt, fill=white] (3,0) circle (2.5pt);
  \draw[black, line width=0.6pt, fill=white] (2,2) circle (2.5pt);
  % spin coupling
  \draw[black, line width=0.6pt, fill=black] (1,0) circle (2.5pt);
  % labels
  \node[above] at (2,2) {$v^0$};
  \node[below] at (1,0) {$v^3$};
  \node[below] at (3,0) {$v^0$};
\end{tikzpicture}

\\
\end{tabular}

\vspace{0pt}

\begin{tabular}{@{}c@{\hspace{2mm}}c@{\hspace{2mm}}c@{}}

\begin{tikzpicture}[x=1cm,y=1cm, line cap=round, scale=0.8, transform shape]
  % interaction
  \draw[phi] (2,1) -- (1,0);
  \draw[A] (2,1) -- (2,2);
  \draw[A] (3,0) -- (2,1);
  % endpoint circles
  \draw[black, line width=0.6pt, fill=white] (1,0) circle (2.5pt);
  \draw[black, line width=0.6pt, fill=white] (3,0) circle (2.5pt);
  \draw[black, line width=0.6pt, fill=white] (2,2) circle (2.5pt);
  % spin coupling
  \draw[black, line width=0.6pt, fill=black] (2,2) circle (2.5pt);
  % labels
  \node[above] at (2,2) {$v^2$};
  \node[below] at (1,0) {$v^0$};
  \node[below] at (3,0) {$v^1$};
\end{tikzpicture}
&
\begin{tikzpicture}[x=1cm,y=1cm, line cap=round, scale=0.8, transform shape]
  % interaction
  \draw[sigma] (2,0.95) -- (2,2);
  \draw[phi] (2,1) -- (1,0);
  \draw[phi] (2,1) -- (3,0);
  % endpoint circles
  \draw[black, line width=0.6pt, fill=white] (1,0) circle (2.5pt);
  \draw[black, line width=0.6pt, fill=white] (3,0) circle (2.5pt);
  \draw[black, line width=0.6pt, fill=white] (2,2) circle (2.5pt);
  % spin coupling
  \draw[black, line width=0.6pt, fill=black] (2,2) circle (2.5pt);
  % labels
  \node[above] at (2,2) {$v^3$};
  \node[below] at (1,0) {$v^0$};
  \node[below] at (3,0) {$v^0$};
\end{tikzpicture}
&
\begin{tikzpicture}[x=1cm,y=1cm, line cap=round, scale=0.8, transform shape]
  % interaction
  \draw[A] (2,2) -- (2,1);
  \draw[phi] (1,0) -- (2,1);
  \draw[phi] (3,0) -- (2,1);
  % endpoint circles
  \draw[black, line width=0.6pt, fill=white] (1,0) circle (2.5pt);
  \draw[black, line width=0.6pt, fill=white] (3,0) circle (2.5pt);
  \draw[black, line width=0.6pt, fill=white] (2,2) circle (2.5pt);
  % spin coupling
  \draw[black, line width=0.6pt, fill=black] (2,2) circle (2.5pt);
  % propagator correction
  \draw[black, line width=0.6pt, fill=white] (2,1) circle (3.5pt);
  \draw[black, line width=0.6pt] ($(2,1)+(-2.2pt,-2.2pt)$) -- ($(2,1)+(2.2pt,2.2pt)$);
  \draw[black, line width=0.6pt] ($(2,1)+(-2.2pt, 2.2pt)$) -- ($(2,1)+(2.2pt,-2.2pt)$);
  % labels
  \node[above] at (2,2) {$v^2$};
  \node[below] at (1,0) {$v^0$};
  \node[below] at (3,0) {$v^0$};
\end{tikzpicture}

\\
\end{tabular}

\end{center}
\caption{Generic worldline diagrams contributing to the NLO spin-orbit interaction. The black and white circles denote spin and mass couplings, respectively. A circled cross denotes a propagator-correction vertex or a time derivative in a cubic bulk vertex.}
\label{fig:NLOgeneric}
\end{figure}
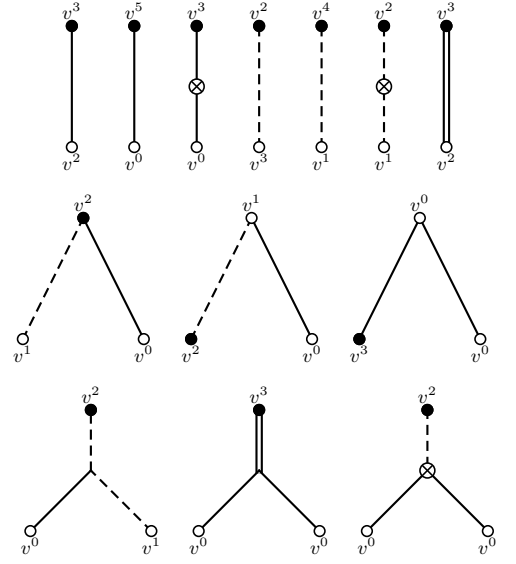

At NLO, the contributions are of order $\mathcal{O}(G)$ and $\mathcal{O}(G^2)$. At $\mathcal{O}(G)$, the only topology is one-graviton exchange, which gives $7$ generic worldline diagrams.
At $\mathcal{O}(G^2)$, two topologies contribute: the ``V'' topology, involving a two-graviton worldline coupling, and the ``Y'' topology, involving a cubic bulk vertex. There are $6$ generic worldline diagrams at $\mathcal{O}(G^2)$: $3$ of ``V'' type and $3$ of ``Y'' type.
The corresponding generic worldline diagrams are shown in Fig.~\ref{fig:NLOgeneric}.
Table~\ref{tab:NLOSO_count} summarizes the diagram count by order and topology, and separates the two-body and three-body contributions.

\begin{table}[t]\caption{Diagram counting in the NLO spin-orbit sector.}
\label{tab:NLOSO_count}
\begin{ruledtabular}
\begin{tabular}{ccccc}
\noalign{\vskip 2pt}
\(G\) order & Topology & Two-body & Three-body & Total 
\\[2pt]\hline\noalign{\vskip 3pt}
$G$ & one-graviton & 7 & 0 & 7 \\[2pt]
\multirow{2}{*}{$G^2$}& ``V'' & 3 & 3 & 6 \\[2pt]
& ``Y'' & 7 & 3 & 10 \\[2pt]
\hline\noalign{\vskip 3pt}
Total & -- & 17 & 6 & 23
\end{tabular}\end{ruledtabular}\end{table}

\subsubsection{NLO two-body diagrams}

The two-body diagrams in the NLO spin-orbit sector are obtained by considering all possible attachments of two source worldlines to the generic worldline diagrams shown in Fig.~\ref{fig:NLOgeneric}. This gives $7$ one-graviton diagrams, $3$ two-graviton diagrams, and $7$ three-graviton diagrams. These diagrams are shown in Fig.~\ref{fig:NLO_SO_diagrams}.
Using the Feynman rules in Table~\ref{tab:Feynman}, we compute all diagrams. They are equivalent to the binary results of~\cite{levi2015spinning}, generalized to $N$ bodies by promoting the binary labels to arbitrary particle labels and summing over all pairs. Terms proportional to $\partial_t S^{ij}$ in the one-graviton diagrams are neglected, since they first contribute at NNLO. The one-graviton diagrams also contain acceleration terms, which are removed in Sec.~\ref{sec:IIIC} by a shift in the position coordinate.

\input{diagrams/NLO_two}

\begin{widetext}
\begin{align}
    L^{(\text{b1})} &= \sum_a\sum_{b\neq a}\frac{3Gm_b}{r_{ab}^2}\,
    \bm{S}_a\!\cdot\bm{v}_a\times\bm{n}_{ab}\,v_b^2, \\[2pt]
   L^{(\text{b2})} &= \sum_a\sum_{b\neq a}\frac{Gm_b}{r_{ab}}\left[
    \bm{S}_a\!\cdot\bm{v}_a\times \bm{n}_{ab}\frac{v_a^2}{r_{ab}}
    +2\bm{S}_a\!\cdot\bm{v}_a\times\bm{a}_a
    \right], \\[2pt]
    L^{(\text{b3})} &= \sum_a\sum_{b\neq a}\frac{Gm_b}{r_{ab}^2}\Big[
    \bm{S}_a\!\cdot\bm{v}_a\times\bm{n}_{ab}\,(\bm{v}_a\!\cdot\bm{v}_b)
    +\bm{S}_a\!\cdot\bm{v}_a\times\bm{v}_b\,(\bm{v}_a\!\cdot\bm{n}_{ab})
    -3\bm{S}_a\!\cdot\bm{v}_a\times\bm{n}_{ab}\,
    (\bm{v}_a\!\cdot\bm{n}_{ab})(\bm{v}_b\!\cdot\bm{n}_{ab})
    \Big]
    \notag\\[2pt]
    &\quad+\sum_a\sum_{b\neq a}\frac{Gm_b}{r_{ab}}\Big[
    \bm{S}_a\!\cdot\bm{v}_b\times\bm{a}_a
    +\bm{S}_a\!\cdot\bm{a}_a\times\bm{n}_{ab}\,(\bm{v}_b\!\cdot\bm{n}_{ab})
    \Big],\\[2pt]
    L^{(\text{b4})} &= -\sum_a\sum_{b\neq a}\frac{Gm_b}{r_{ab}^2}\,
    \bm{S}_a\!\cdot \bm{v}_b\times \bm{n}_{ab}\,v_b^2, \\[2pt]
    L^{(\text{b5})} &= -\sum_a\sum_{b\neq a}\frac{Gm_b}{r_{ab}^2}\Big[
    3\bm{S}_a\!\cdot\bm{v}_a\times\bm{n}_{ab}\,(\bm{v}_a\!\cdot\bm{v}_b)
    +\bm{S}_a\!\cdot\bm{v}_a\times\bm{v}_b\,(\bm{v}_a\!\cdot\bm{n}_{ab})
    \Big]
    \notag\\[2pt]
    &\quad+\sum_a\sum_{b\neq a}\frac{4Gm_b}{r_{ab}}
    \bm{S}_a\!\cdot\bm{a}_a\times\bm{v}_b, \\[2pt]
    L^{(\text{b6})}
    &= \sum_a\sum_{b\neq a}\frac{Gm_b}{r_{ab}^{2}}
    \Big[\!
    -\bm{S}_a \!\cdot\bm{v}_b \times \bm{n}_{ab}(\bm{v}_a \!\cdot \bm{v}_b)
    + \bm{S}_a \!\cdot\bm{v}_a \times \bm{v}_b(\bm{v}_b \!\cdot \bm{n}_{ab})
    + 3 \bm{S}_a \!\cdot\bm{v}_b \times \bm{n}_{ab}(\bm{v}_a \!\cdot \bm{n}_{ab})(\bm{v}_b \!\cdot \bm{n}_{ab})
    \Big]
    \notag \\
    &\quad+ \sum_a\sum_{b\neq a}\frac{Gm_b}{r_{ab}}
    \Big[\bm{S}_a \!\cdot \bm{v}_a \times \bm{a}_b
    +\bm{S}_a\!\cdot \bm{a}_b\times\bm{n}_{ab}\,(\bm{v}_a\!\cdot\bm{n}_{ab})
    \Big], \\[2pt]
    L^{(\text{b7})}
    &= -\sum_a\sum_{b\neq a}\frac{2Gm_b}{r_{ab}^{2}} 
    \Big[\bm{S}_a \!\cdot
    \bm{v}_a \times \bm{n}_{ab}\,v_b^{2}
    - \bm{S}_a \!\cdot\bm{v}_b \times \bm{n}_{ab}(\bm{v}_a \!\cdot \bm{v}_b)
    \Big],\\[2pt]
    L^{(\text{c1})} &= \sum_a\sum_{b\neq a}\frac{8G^2m_b^2}{r_{ab}^3}\bm{S}_a\!\cdot\bm{v}_b\times \bm{n}_{ab},\\[2pt]
    L^{(\text{c2})} &= \sum_a\sum_{b\neq a}\frac{2 G^{2} m_a m_b}{r_{ab}^{3}} \bm{S}_a \!\cdot \bm{v}_b \times \bm{n}_{ab},\\[2pt]
    L^{(\text{c3})} &= -\sum_a\sum_{b\neq a}\frac{2G^2m_a m_b}{r_{ab}^3}\bm{S}_a\!\cdot\bm{v}_a\times\bm{n}_{ab},\\[2pt]
    L^{(\text{d1})} &= -\sum_a\sum_{b\neq a}\frac{8 G^{2} m_b^{2}}{r_{ab}^{3}} \bm{S}_a \!\cdot \bm{v}_b \times \bm{n}_{ab}
    ,\\[2pt]
    L^{(\text{d2})} &= -\sum_a\sum_{b\neq a}\frac{2 G^{2} m_a m_b}{r_{ab}^{3}} \bm{S}_a \!\cdot \bm{v}_b \times \bm{n}_{ab}
    ,\\[2pt]
    L^{(\text{d3})} &= \sum_a\sum_{b\neq a}\frac{2 G^{2} m_a m_b}{r_{ab}^{3}} \bm{S}_a \!\cdot \bm{v}_a \times \bm{n}_{ab}
    ,\\[2pt]
    L^{(\text{d4})} &= -\sum_a\sum_{b\neq a}\frac{G^{2} m_b^{2}}{2 r_{ab}^{3}} \bm{S}_a \!\cdot \bm{v}_a \times \bm{n}_{ab}
    ,\\[2pt]
    L^{(\text{d5})} &= -\sum_a\sum_{b\neq a}\frac{G^{2} m_a m_b}{2 r_{ab}^{3}} \bm{S}_a \!\cdot \bm{v}_a \times \bm{n}_{ab}
    ,\\[2pt]
    L^{(\text{d6})} &= \sum_a\sum_{b\neq a}\frac{G^{2} m_b^{2}}{2 r_{ab}^{3}} \bm{S}_a \!\cdot \bm{v}_b \times \bm{n}_{ab}
    ,\\[2pt]
    L^{(\text{d7})} &= \sum_a\sum_{b\neq a}\frac{G^{2} m_a m_b}{2 r_{ab}^{3}} \bm{S}_a \!\cdot \bm{v}_a \times \bm{n}_{ab}
    .
\end{align}
\end{widetext}

\subsubsection{NLO three-body diagrams}

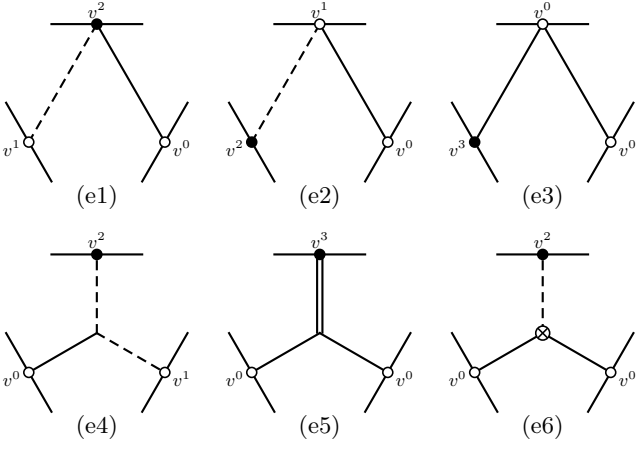
\begin{figure}[t!]

\begin{center}

\begin{tabular}{@{}c@{}}

\begin{tabular}{@{}c@{\hspace{3mm}}c@{\hspace{3mm}}c@{}}
\begin{tikzpicture}[x=1cm,y=1cm, line cap=round, scale=0.75, transform shape]
  % interaction
  \draw[A] (-1.2,-0.6928) -- (0,1.3856);
  \draw[phi] (0,1.3856) -- (1.2,-0.6928);
  % black lines
  \draw[worldline] (-0.8,1.3856) -- (0.8,1.3856);
  \draw[worldline] (-1.6,0) -- (-0.8,-1.3856);
  \draw[worldline] (0.8,-1.3856) -- (1.6,0);
  % worldline vertices
  \draw[black, line width=0.6pt, fill=white] (-1.2,-0.6928) circle (2.5pt);
  \draw[black, line width=0.6pt, fill=white] (1.2,-0.6928) circle (2.5pt);
  % spin coupling
  \draw[black, line width=0.6pt, fill=black] (0,1.3856) circle (2.5pt);
  % labels
  \node[above] at (0,1.3856) {$v^2$};
  \node[below] at (-1.5,-0.55) {$v^1$};
  \node[below] at (1.5,-0.55) {$v^0$};
\end{tikzpicture}
&
\begin{tikzpicture}[x=1cm,y=1cm, line cap=round, scale=0.75, transform shape]
  % interaction
  \draw[A] (-1.2,-0.6928) -- (0,1.3856);
  \draw[phi] (1.2,-0.6928) -- (0,1.3856);
  % black lines
  \draw[worldline] (-0.8,1.3856) -- (0.8,1.3856);
  \draw[worldline] (-1.6,0) -- (-0.8,-1.3856);
  \draw[worldline] (0.8,-1.3856) -- (1.6,0);
  % worldline vertices
  \draw[black, line width=0.6pt, fill=white] (0,1.3856) circle (2.5pt);
  \draw[black, line width=0.6pt, fill=white] (1.2,-0.6928) circle (2.5pt);
  % spin coupling
  \draw[black, line width=0.6pt, fill=black] (-1.2,-0.6928) circle (2.5pt);
  % labels
  \node[above] at (0,1.3856) {$v^1$};
  \node[below] at (-1.5,-0.55) {$v^2$};
  \node[below] at (1.5,-0.55) {$v^0$};
\end{tikzpicture}
&
\begin{tikzpicture}[x=1cm,y=1cm, line cap=round, scale=0.75, transform shape]
  % interaction
  \draw[phi] (-1.2,-0.6928) -- (0,1.3856);
  \draw[phi] (1.2,-0.6928) -- (0,1.3856);
  % black lines
  \draw[worldline] (-0.8,1.3856) -- (0.8,1.3856);
  \draw[worldline] (-1.6,0) -- (-0.8,-1.3856);
  \draw[worldline] (0.8,-1.3856) -- (1.6,0);
  % worldline vertices
  \draw[black, line width=0.6pt, fill=white] (0,1.3856) circle (2.5pt);
  \draw[black, line width=0.6pt, fill=white] (1.2,-0.6928) circle (2.5pt);
  % spin coupling
  \draw[black, line width=0.6pt, fill=black] (-1.2,-0.6928) circle (2.5pt);
  % labels
  \node[above] at (0,1.3856) {$v^0$};
  \node[below] at (-1.5,-0.55) {$v^3$};
  \node[below] at (1.5,-0.55) {$v^0$};
\end{tikzpicture}
\\[-1mm]
(e1) & (e2) & (e3)
\\[2mm]

\begin{tikzpicture}[x=1cm,y=1cm, line cap=round, scale=0.75, transform shape]
  % interaction
  \draw[phi] (0,0) -- (-1.2,-0.6928);
  \draw[A] (0,0) -- (0,1.3856);
  \draw[A] (0,0) -- (1.2,-0.6928);
  % black lines
  \draw[worldline] (-0.8,1.3856) -- (0.8,1.3856);
  \draw[worldline] (-1.6,0) -- (-0.8,-1.3856);
  \draw[worldline] (0.8,-1.3856) -- (1.6,0);
  % worldline vertices
  \draw[black, line width=0.6pt, fill=white] (-1.2,-0.6928) circle (2.5pt);
  \draw[black, line width=0.6pt, fill=white] (1.2,-0.6928) circle (2.5pt);
  % spin coupling
  \draw[black, line width=0.6pt, fill=black] (0,1.3856) circle (2.5pt);
  % labels
  \node[above] at (0,1.3856) {$v^2$};
  \node[below] at (-1.5,-0.55) {$v^0$};
  \node[below] at (1.5,-0.55) {$v^1$};
\end{tikzpicture}
&
\begin{tikzpicture}[x=1cm,y=1cm, line cap=round, scale=0.75, transform shape]
  % interaction
  \draw[sigma] (0,-0.01) -- (0,1.3856);
  \draw[phi] (0,0) -- (-1.2,-0.6928);
  \draw[phi] (0,0) -- (1.2,-0.6928);
  % black lines
  \draw[worldline] (-0.8,1.3856) -- (0.8,1.3856);
  \draw[worldline] (-1.6,0) -- (-0.8,-1.3856);
  \draw[worldline] (0.8,-1.3856) -- (1.6,0);
  % worldline vertices
  \draw[black, line width=0.6pt, fill=white] (-1.2,-0.6928) circle (2.5pt);
  \draw[black, line width=0.6pt, fill=white] (1.2,-0.6928) circle (2.5pt);
  % spin coupling
  \draw[black, line width=0.6pt, fill=black] (0,1.3856) circle (2.5pt);
  % labels
  \node[above] at (0,1.3856) {$v^3$};
  \node[below] at (-1.5,-0.55) {$v^0$};
  \node[below] at (1.5,-0.55) {$v^0$};
\end{tikzpicture}
&
\begin{tikzpicture}[x=1cm,y=1cm, line cap=round, scale=0.75, transform shape]
  % interaction
  \draw[A] (0,0) -- (0,1.3856);
  \draw[phi] (0,0) -- (-1.2,-0.6928);
  \draw[phi] (1.2,-0.6928) -- (0,0);
  % black lines
  \draw[worldline] (-0.8,1.3856) -- (0.8,1.3856);
  \draw[worldline] (-1.6,0) -- (-0.8,-1.3856);
  \draw[worldline] (0.8,-1.3856) -- (1.6,0);
  % worldline vertices
  \draw[black, line width=0.6pt, fill=white] (-1.2,-0.6928) circle (2.5pt);
  \draw[black, line width=0.6pt, fill=white] (1.2,-0.6928) circle (2.5pt);
  % spin coupling
  \draw[black, line width=0.6pt, fill=black] (0,1.3856) circle (2.5pt);
  % propagator correction
  \draw[black, line width=0.6pt, fill=white] (0,0) circle (3.5pt);
  \draw[black, line width=0.6pt] ($(0,0)+(-2.2pt,-2.2pt)$) -- ($(0,0)+(2.2pt,2.2pt)$);
  \draw[black, line width=0.6pt] ($(0,0)+(-2.2pt, 2.2pt)$) -- ($(0,0)+(2.2pt,-2.2pt)$);
  % labels
  \node[above] at (0,1.3856) {$v^2$};
  \node[below] at (-1.5,-0.55) {$v^0$};
  \node[below] at (1.5,-0.55) {$v^0$};
\end{tikzpicture}
\\[-1mm]
(e4) & (e5) & (e6)
\end{tabular}

\end{tabular}

\end{center}

\caption{New three-body diagrams contributing to the NLO spin-orbit sector. The top row shows the two-graviton diagrams, and the bottom row shows the three-graviton diagrams. The black and white circles denote spin and mass couplings, respectively. The circled cross denotes a time derivative in the cubic bulk vertex. All diagrams have symmetry factor $1$.}
\label{fig:NLOthree}
\end{figure}
Attaching $3$ source worldlines to the two-graviton and three-graviton generic worldline diagrams in Fig.~\ref{fig:NLOgeneric} gives the $6$ three-body diagrams shown in Fig.~\ref{fig:NLOthree}. They correspond to the new contributions specific to the $N$-body problem. 
Using the Feynman rules in Table~\ref{tab:Feynman}, we obtain
\begin{widetext}
\begin{align}
    L^{(\text{e1})} &= 
    \sum_a\sum_{b\neq a}\sum_{c\neq a,b}\frac{8G^2 m_b m_c}{r_{ab}^2r_{ac}}\bm{S}_a\!\cdot \bm{v}_b\times\bm{n}_{ab} ,\\[2pt]
    L^{(\text{e2})} &= 
    \sum_a\sum_{b\neq a}\sum_{c\neq a,b} \frac{2G^2m_bm_c}{r_{ab}^2r_{bc}}\bm{S}_a\!\cdot \bm{v}_b\times \bm{n}_{ab} ,\\[2pt]
    L^{(\text{e3})} &=  
    -\sum_a\sum_{b\neq a}\sum_{c\neq a,b}\frac{2G^2m_b m_c}{r_{ab}^2r_{bc}}\bm{S}_a\!\cdot\bm{v}_a\times\bm{n}_{ab}, \\[2pt]
    L^{(\text{e4})} &= 
    -\sum_a\sum_{b\neq a}\sum_{c\neq a,b} 8G^2m_bm_c \Bigg\{
    \frac{1}{r_{ab}s_{abc}^2}
    \big[
    \bm{S}_a \!\cdot \bm{v}_c\!\times\!\bm{n}_{bc}
    +\bm{S}_a \!\cdot \bm{v}_c\!\times\!\bm{n}_{ac}
    -(\bm{v}_c\!\cdot\bm{n}_{ab})
    \left(
    \bm{S}_a \!\cdot \bm{n}_{ab}\!\times\!\bm{n}_{ac}
    +\bm{S}_a \!\cdot \bm{n}_{ab}\!\times\!\bm{n}_{bc}
    \right)
    \big]
    \notag \\[2pt]
    &\quad\quad+
    \frac{1}{r_{ac}s_{abc}^2}
    \big[
    \bm{S}_a \!\cdot \bm{v}_c\!\times\!\bm{n}_{bc}
    -\bm{S}_a \!\cdot \bm{v}_c\!\times\!\bm{n}_{ab}-(\bm{v}_c\!\cdot\bm{n}_{ac})
    \left(
    \bm{S}_a \!\cdot \bm{n}_{ab}\!\times\!\bm{n}_{ac}+
    \bm{S}_a \!\cdot \bm{n}_{ac}\!\times\!\bm{n}_{bc}
    \right)
    \big]
    \notag \\[2pt]
    &\quad\quad-
    \frac{2}{s_{abc}^3}
    \big[
    \left(\bm{v}_c\!\cdot\bm{n}_{ab}+\bm{v}_c\!\cdot\bm{n}_{ac}\right)
    \left(
    \bm{S}_a \!\cdot \bm{n}_{ab}\!\times\!\bm{n}_{ac}
    +\bm{S}_a \!\cdot \bm{n}_{ab}\!\times\!\bm{n}_{bc}
    +\bm{S}_a \!\cdot \bm{n}_{ac}\!\times\!\bm{n}_{bc}
    \right)
    \big]\notag \\[2pt]
    &\quad\quad
    +\frac{r_{ac}+r_{bc}}{2r_{ab}^{2}r_{ac}r_{bc}}\,\bm{S}_a \!\cdot \bm{v}_b\!\times\!\bm{n}_{ab}
    +\frac{r_{bc}-r_{ac}}{2r_{ab}^{2}r_{bc}r_{ac}}\,\bm{S}_a \!\cdot \bm{v}_c\!\times\!\bm{n}_{ab}
    \Bigg\},
    \\[2pt]
    L^{(\text{e5})} &= \sum_a\sum_{b\neq a}\sum_{c\neq a,b} G^2m_bm_c \Bigg\{
    \frac{1}{r_{ab}s_{abc}^{2}}
    \big[
    \bm{S}_a \!\cdot \bm{v}_a\!\times\!\bm{n}_{ac}
    +\bm{S}_a \!\cdot \bm{v}_a\!\times\!\bm{n}_{bc}
    -(\bm{v}_a\!\cdot\bm{n}_{ab})
    \left(
    \bm{S}_a \!\cdot \bm{n}_{ab}\!\times\!\bm{n}_{ac}
    +\bm{S}_a \!\cdot \bm{n}_{ab}\!\times\!\bm{n}_{bc}
    \right)
    \big]
    \notag\\[2pt]
    &\quad\quad+
    \frac{1}{r_{ac}s_{abc}^{2}}
    \big[
    \bm{S}_a \!\cdot \bm{v}_a\!\times\!\bm{n}_{ab}
    -\bm{S}_a \!\cdot \bm{v}_a\!\times\!\bm{n}_{bc}
    +(\bm{v}_a\!\cdot\bm{n}_{ac})
    \left(
    \bm{S}_a \!\cdot \bm{n}_{ab}\!\times\!\bm{n}_{ac}
    +\bm{S}_a \!\cdot \bm{n}_{ac}\!\times\!\bm{n}_{bc}
    \right)
    \big]
    \notag\\[2pt]
    &\quad\quad+
    \frac{2}{r_{bc}s_{abc}^{2}}
    \big[
    \bm{S}_a \!\cdot \bm{v}_a\!\times\!\bm{n}_{ab}
    +\bm{S}_a \!\cdot \bm{v}_a\!\times\!\bm{n}_{ac} +(\bm{v}_a\!\cdot\bm{n}_{bc})
    \left(
    \bm{S}_a \!\cdot \bm{n}_{ab}\!\times\!\bm{n}_{bc}
    +\bm{S}_a \!\cdot \bm{n}_{ac}\!\times\!\bm{n}_{bc}
    \right)
    \big]
    \notag\\[2pt]
    &\quad\quad-
    \frac{2}{s_{abc}^{3}}
    \big[
    \left(
    \bm{v}_a\!\cdot\bm{n}_{ab}
    -\bm{v}_a\!\cdot\bm{n}_{ac}
    -2\bm{v}_a\!\cdot\bm{n}_{bc}
    \right)
    \left(
    \bm{S}_a \!\cdot \bm{n}_{ab}\!\times\!\bm{n}_{ac}
    +\bm{S}_a \!\cdot \bm{n}_{ab}\!\times\!\bm{n}_{bc}
    +\bm{S}_a \!\cdot \bm{n}_{ac}\!\times\!\bm{n}_{bc}
    \right)
    \big]
    \Bigg\},
    \\[2pt]
    L^{(\text{e6})} &=  -\sum_a\sum_{b\neq a}\sum_{c\neq a,b} G^2m_bm_c \Bigg\{
    \frac{1}{r_{ab}s_{abc}^{2}}
    \big[
    \bm{S}_a \!\cdot \bm{v}_b\!\times\!\bm{n}_{ac}
    +\bm{S}_a \!\cdot \bm{v}_b\!\times\!\bm{n}_{bc}
    -\left(\bm{v}_{b}\!\cdot\bm{n}_{ab}\right)
    \left(
    \bm{S}_a \!\cdot \bm{n}_{ab}\!\times\!\bm{n}_{ac}
    +\bm{S}_a \!\cdot \bm{n}_{ab}\!\times\!\bm{n}_{bc}
    \right)
    \big]
    \notag\\[2pt]
    &\quad\quad+
    \frac{1}{r_{ac}s_{abc}^{2}}
    \big[
    \bm{S}_a \!\cdot \bm{v}_c\!\times\!\bm{n}_{ab}
    -\bm{S}_a \!\cdot \bm{v}_c\!\times\!\bm{n}_{bc}
    +\left(\bm{v}_{c}\!\cdot\bm{n}_{ac}\right)
    \left(
    \bm{S}_a \!\cdot \bm{n}_{ab}\!\times\!\bm{n}_{ac}
    +\bm{S}_a \!\cdot \bm{n}_{ac}\!\times\!\bm{n}_{bc}
    \right)
    \big]
    \notag\\[2pt]
    &\quad\quad+
    \frac{1}{r_{bc}s_{abc}^{2}}
    \big[
    \bm{S}_a \!\cdot \bm{v}_b\!\times\!\bm{n}_{ab}
    +\bm{S}_a \!\cdot \bm{v}_b\!\times\!\bm{n}_{ac}
    +\bm{S}_a \!\cdot \bm{v}_c\!\times\!\bm{n}_{ab}
    +\bm{S}_a \!\cdot \bm{v}_c\!\times\!\bm{n}_{ac}
    \notag\\[2pt]
    &\quad\quad\quad+
    \left(\bm{v}_{b}\!\cdot\bm{n}_{bc}+\bm{v}_{c}\!\cdot\bm{n}_{bc}\right)
    \left(
    \bm{S}_a \!\cdot \bm{n}_{ab}\!\times\!\bm{n}_{bc}
    +\bm{S}_a \!\cdot \bm{n}_{ac}\!\times\!\bm{n}_{bc}
    \right)
    \big]
    \notag\\[2pt]
    &\quad\quad+
    \frac{2}{s_{abc}^{3}}
    \big[
    \left(
    \bm{v}_{c}\!\cdot (\bm{n}_{ac}+\bm{n}_{bc})
    -\bm{v}_{b}\!\cdot (\bm{n}_{ab}-\bm{n}_{bc})
    \right)
    \left(
    \bm{S}_a \!\cdot \bm{n}_{ab}\!\times\!\bm{n}_{ac}
    +\bm{S}_a \!\cdot \bm{n}_{ab}\!\times\!\bm{n}_{bc}
    +\bm{S}_a \!\cdot \bm{n}_{ac}\!\times\!\bm{n}_{bc}
    \right)
    \big]
    \Bigg\}. 
\end{align}
\end{widetext}
Here $s_{abc}\equiv r_{ab}+r_{bc}+r_{ac}$. In the computation of three-graviton diagrams, we used the Fourier representation of the triangle potential~\cite{will2025},
\begin{equation}
    \frac{1-\ln s_{abc}}{16\pi^2}
    =
    \int_{\bm{p},\bm{q}}
    \frac{
    e^{i\bm{p}\cdot\bm{r}_{ab}}
    e^{-i\bm{q}\cdot\bm{r}_{bc}}
    }
    {\bm{p}^2\bm{q}^2(\bm{p}+\bm{q})^2},
\end{equation}
which we used to express the momentum integrals as derivatives of $\ln s_{abc}$.

\subsection{NLO spin-orbit potential}
\label{sec:IIIC}

Adding the kinetic term in Eq.~\eqref{lkin} to the contributions from all Feynman diagrams, we obtain the LO and NLO spin-orbit potentials,
\begin{widetext}
\begin{align}
    V^{\rm LO}_{\rm SO} &= 
    -\sum_a \frac{1}{2}\bm{S}_a\!\cdot \bm{v}_a\times \bm{a}_a 
    +\sum_a\sum_{b\neq a}\frac{2Gm_b}{r_{ab}^2}\bm{S}_a\!\cdot (\bm{v}_b-\bm{v}_a)\times\bm{n}_{ab}, 
    \\[2pt]
    V_{\rm SO}^{\rm NLO}
    &=
\sum_a
\frac{3}{8}v_a^2\,
\bm{S}_a\!\cdot\bm{a}_a\times\bm{v}_a
\notag\\[2pt]
&\quad
-\sum_a\sum_{b\neq a}
\frac{Gm_b}{r_{ab}^{2}}
\bigg[
\bm{S}_a\!\cdot\bm{v}_a\times\bm{n}_{ab}
\bigg(
v_a^2
-2(\bm{v}_a\!\cdot\bm{v}_b)
+v_b^2
-3(\bm{v}_a\!\cdot\bm{n}_{ab})
(\bm{v}_b\!\cdot\bm{n}_{ab})
\bigg)
\notag\\[2pt]
&\quad\quad
+\bm{S}_a\!\cdot\bm{v}_b\times\bm{n}_{ab}
\bigg(
(\bm{v}_a\!\cdot\bm{v}_b)
-v_b^2
+3(\bm{v}_a\!\cdot\bm{n}_{ab})
(\bm{v}_b\!\cdot\bm{n}_{ab})
\bigg)
+\bm{S}_a\!\cdot\bm{v}_a\times\bm{v}_b\,
(\bm{v}_b\!\cdot\bm{n}_{ab})
\bigg]
\notag\\[2pt]
&\quad
+\sum_a\sum_{b\neq a}
\frac{G^2m_b^2}{2r_{ab}^3}\,
\bm{S}_a\!\cdot(\bm{v}_a-\bm{v}_b)\times\bm{n}_{ab}
\notag\\[2pt]
&\quad
+\sum_a\sum_{b\neq a}
\frac{Gm_b}{r_{ab}}
\bigg[
\bm{S}_a\!\cdot\bm{a}_a\times
\big(
2\bm{v}_a-3\bm{v}_b
-\bm{n}_{ab}(\bm{v}_b\!\cdot\bm{n}_{ab})
\big)
+\bm{S}_a\!\cdot\bm{a}_b\times
\big(
\bm{v}_a-\bm{n}_{ab}(\bm{v}_a\!\cdot\bm{n}_{ab})
\big)
\bigg]
\notag\\[2pt]
&\quad
+\sum_a\sum_{b\neq a}\sum_{c\neq a,b}
G^2m_bm_c
\Bigg\{
\frac{2}{r_{ab}^2}\,
\bm{S}_a\!\cdot
\bigg(
\frac{\bm{v}_a+\bm{v}_b-2\bm{v}_c}{r_{bc}}
+\frac{2(\bm{v}_c-\bm{v}_b)}{r_{ac}}
\bigg)\times\bm{n}_{ab}
\notag\\[2pt]
&\quad\quad
+\frac{1}{r_{ab}s_{abc}^{2}}
\Big[
\bm{S}_a\!\cdot
(8\bm{v}_c-3\bm{v}_a+\bm{v}_b)\!
\times(\bm{n}_{ac}+\bm{n}_{bc})
+\big(
3(\bm{v}_a\!\cdot\bm{n}_{ab})
-(\bm{v}_b\!\cdot\bm{n}_{ab})
-8(\bm{v}_c\!\cdot\bm{n}_{ab})
\big)\,
\bm{S}_a\!\cdot\bm{n}_{ab}\times
(\bm{n}_{ac}+\bm{n}_{bc})
\Big]
\notag\\[2pt]
&\quad\quad
+\frac{1}{r_{ac}s_{abc}^{2}}
\Big[
\bm{S}_a\!\cdot
(\bm{v}_a-7\bm{v}_c)
\times(\bm{n}_{ab}-\bm{n}_{bc})
+\big(
(\bm{v}_a\!\cdot\bm{n}_{ac})
-7(\bm{v}_c\!\cdot\bm{n}_{ac})
\big)\,
\bm{S}_a\!\cdot
(\bm{n}_{ab}-\bm{n}_{bc})\times\bm{n}_{ac}
\Big]
\notag\\[2pt]
&\quad\quad
+\frac{2}{r_{bc}s_{abc}^{2}}
\Big[\!
-2\bm{S}_a\!\cdot\bm{v}_a\times\bm{n}_{ab}
+\bm{S}_a\!\cdot\bm{v}_b\times
(\bm{n}_{ab}+\bm{n}_{ac})
+\big(
(\bm{v}_b\!\cdot\bm{n}_{bc})
+(\bm{v}_c\!\cdot\bm{n}_{bc})
-2(\bm{v}_a\!\cdot\bm{n}_{bc})
\big)\,
\bm{S}_a\!\cdot\bm{n}_{ab}\times\bm{n}_{bc}
\Big]
\notag\\[2pt]
&\quad\quad
+\frac{2}{s_{abc}^{3}}
\Big[
(\bm{v}_a\!\cdot\bm{n}_{ab})
-(\bm{v}_a\!\cdot\bm{n}_{ac})
-2(\bm{v}_a\!\cdot\bm{n}_{bc})
+3(\bm{v}_b\!\cdot\bm{n}_{ab})
+4(\bm{v}_b\!\cdot\bm{n}_{ac})
+(\bm{v}_b\!\cdot\bm{n}_{bc})
\notag\\[2pt]
&\quad\quad\quad
-4(\bm{v}_c\!\cdot\bm{n}_{ab})
-3(\bm{v}_c\!\cdot\bm{n}_{ac})
+(\bm{v}_c\!\cdot\bm{n}_{bc})
\Big]
\left(
\bm{S}_a\!\cdot\bm{n}_{ab}\times\bm{n}_{ac}
+2\bm{S}_a\!\cdot\bm{n}_{ab}\times\bm{n}_{bc}
\right)
\Bigg\}.
\end{align}
\end{widetext}
The two-body parts of $V_{\rm SO}^{\rm LO}$ and $V_{\rm SO}^{\rm NLO}$ reproduce the binary generalized-canonical-gauge results of Eqs.~(6.3) and (6.18) of~\cite{levi2015spinning}, respectively, after promoting the binary labels to arbitrary body labels and summing over pairs.
In the NLO expression, we omit time derivatives of the spin.
The remaining terms are genuine three-body contributions obtained from the 6 new three-body diagrams (e1)--(e6).

To obtain the NLO spin-orbit potential in acceleration-free form, we remove the acceleration-dependent terms. This is done by the following position shift~\cite{levi2015spinning}, which is equivalent to using 1PN equations of motion:
\begin{widetext}
\begin{align}
\bm{x}_a \to \bm{x}_a
&
+\frac{1}{2m_a}\bm{S}_a\times \bm{v}_a
+ \frac{1}{8m_a}\,\bm{S}_a \times \bm{v}_a\, v_a^2 \nonumber\\
&+ \sum_{b\neq a}\frac{Gm_b}{m_a r_{ab}}
\left(
\frac{1}{2}\,\bm{S}_a \times \bm{v}_a
-3\,\bm{S}_a \times \bm{v}_b
-\bm{S}_a \times \bm{n}_{ab}(\bm{v}_b\!\cdot\bm{n}_{ab})
\right) \nonumber\\
&+ \sum_{b\neq a}\frac{G}{r_{ab}}
\left(
\frac{11}{4}\,\bm{S}_b \times \bm{v}_b
-\bm{S}_b \times \bm{n}_{ab}(\bm{v}_b\!\cdot\bm{n}_{ab})
+\frac{1}{4}\,\bm{n}_{ab}\,\bm{S}_b\!\cdot\bm{v}_b\times\bm{n}_{ab}
\right).
\end{align}
\end{widetext}
Under a position shift, the potential changes by
\begin{equation}
    \Delta V=-\sum_a \Delta \bm{x}_a \!\cdot \left(\frac{\partial L}{\partial \bm{x}_a}-\frac{d}{dt}\frac{\partial L}{\partial \bm{v}_a}\right)+\mathcal{O}\!\left((\Delta \bm{x}_a)^2\right).
\end{equation}
Since $\Delta\bm{x}_a$ is linear in spin, the shift only affects spin-dependent terms. Terms of order $\mathcal{O}\!\left((\Delta\bm{x}_a)^2\right)$ are higher order in spin and are therefore neglected.
To the required order, we evaluate $\Delta V$ using $L=L_{\text{N}}+L_{\text{EIH}}$ for the leading $\mathcal{O}(Sv)$ part of $\Delta\bm{x}_a$, and using only $L=L_{\text{N}}$ for the higher-order parts of $\Delta\bm{x}_a$. 
Here $L_{\text{N}}$ is the Newtonian Lagrangian,
\begin{equation}
    L_{\text{N}}=\frac{1}{2}\sum_{a}m_a v_a^2+\sum_{b>a}\frac{Gm_a m_b}{r_{ab}},
\end{equation}
and $L_{\text{EIH}}$ is the Einstein-Infeld-Hoffmann Lagrangian,
\begin{align}
L_{\rm EIH}
&=
\frac{1}{8}\sum_{a} m_a v_a^4 
+\sum_{b>a}
\frac{G m_a m_b}{2r_{ab}}
\Big[\,
3v_a^2+3v_b^2
\notag\\[2pt]
&\qquad
-7\bm{v}_a\!\cdot\bm{v}_b
-(\bm{v}_a\!\cdot\bm{n}_{ab})(\bm{v}_b\!\cdot\bm{n}_{ab})
\Big]
\notag\\[2pt]
&\quad-
\sum_{a}
\sum_{ b\neq a}
\sum_{ c\neq a}
\frac{G^2m_a m_b m_c}{2r_{ab}r_{ac}} .
\end{align}
The resulting shift of the potential is
\begin{equation}
    \Delta V = \Delta V_{\rm LO}+\Delta V_{\rm NLO},
\end{equation}
where
\begin{widetext}
\begin{align}
    \Delta V_{\rm LO} 
    &=\sum_a \frac{1}{2}\bm{S}_a\!\cdot \bm{v}_a\times\bm{a}_a + \sum_a \sum_{b\neq a}\frac{Gm_b}{2 r_{ab}^2}\bm{S}_a\!\cdot \bm{v}_a\times\bm{n}_{ab}
    ,\\[2pt]
    \Delta V_{\rm NLO} 
    &=
    \sum_a
    \frac{3}{8}v_a^2\,
    \bm{S}_a\!\cdot\bm{v}_a\times\bm{a}_a
    \notag\\[2pt]
    &\quad +
    \sum_a\sum_{b\neq a}
    \frac{Gm_b}{r_{ab}^{2}}
    \bigg[\,
    \frac{7}{8}\bm{S}_a\!\cdot\bm{v}_a\times\bm{n}_{ab}\,v_a^2
    -2\bm{S}_a\!\cdot\bm{v}_a\times\bm{n}_{ab}\,
    (\bm{v}_a\!\cdot\bm{v}_b)
    +\bm{S}_a\!\cdot\bm{v}_a\times\bm{n}_{ab}\,v_b^2
    \notag\\[2pt]
    &\quad\quad
    -\frac{3}{4}\bm{S}_a\!\cdot\bm{v}_a\times\bm{n}_{ab}\,
    (\bm{v}_b\!\cdot\bm{n}_{ab})^2
    +2\bm{S}_a\!\cdot\bm{v}_a\times\bm{v}_b\,
    (\bm{v}_a\!\cdot\bm{n}_{ab})
    -\frac{3}{2}\bm{S}_a\!\cdot\bm{v}_a\times\bm{v}_b\,
    (\bm{v}_b\!\cdot\bm{n}_{ab})
    \bigg]
    \notag\\[2pt]
    &\quad
    -\sum_a\sum_{b\neq a}
    \frac{G^2m_b}{r_{ab}^{3}}
    \bigg[\,
    \frac{7}{2}m_a\,
    \bm{S}_a\!\cdot\bm{v}_a\times\bm{n}_{ab}
    +3m_b\,
    \bm{S}_a\!\cdot\bm{v}_b\times\bm{n}_{ab}
    \bigg]
    \notag\\[2pt]
    &\quad
    +\sum_a\sum_{b\neq a}
    \frac{Gm_b}{r_{ab}}
    \Big[
    \bm{S}_a\!\cdot\bm{a}_b\times\bm{n}_{ab}\,
    (\bm{v}_a\!\cdot\bm{n}_{ab})
    +\bm{S}_a\!\cdot\bm{a}_a\times\bm{n}_{ab}\,
    (\bm{v}_b\!\cdot\bm{n}_{ab})
    \notag\\[2pt]
    &\quad\quad
    -2\bm{S}_a\!\cdot\bm{a}_a\times\bm{v}_a
    -\bm{S}_a\!\cdot\bm{a}_b\times\bm{v}_a
    +3\bm{S}_a\!\cdot\bm{a}_a\times\bm{v}_b
    \Big]
    \notag\\[2pt]
    &\quad +
    \sum_a\sum_{b\neq a}\sum_{c\neq a,b}
    G^2m_bm_c
    \Bigg\{
    \frac{1}{r_{ab}r_{bc}^{2}}
    \bigg[\!
    -\bm{S}_a\!\cdot\bm{n}_{ab}\times\bm{n}_{bc}\,
    (\bm{v}_a\!\cdot\bm{n}_{ab})
    +\frac{11}{4}\bm{S}_a\!\cdot\bm{v}_a\times\bm{n}_{bc}
    +\frac{1}{4}
    (\bm{n}_{ab}\!\cdot\bm{n}_{bc})\,
    \bm{S}_a\!\cdot\bm{v}_a\times\bm{n}_{ab}
    \bigg]
    \notag\\[2pt]
    &\quad\quad
    -\frac{1}{2r_{ab}^{2}r_{bc}}\,
    \bm{S}_a\!\cdot\bm{v}_a\times\bm{n}_{ab}
    -\frac{1}{r_{ab}r_{ac}^{2}}
    \Big[
    \bm{S}_a\!\cdot\bm{n}_{ab}\times\bm{n}_{ac}\,
    (\bm{v}_b\!\cdot\bm{n}_{ab})
    +3\bm{S}_a\!\cdot\bm{v}_b\times\bm{n}_{ac}
    \Big]
    \Bigg\}.
\end{align}
\end{widetext}

After the position shift, the LO and NLO spin-orbit potentials become
\begin{widetext}
\begin{align}
    V_{\rm SO}^{\rm LO} &= \sum_a\sum_{b\neq a}\frac{Gm_b}{r_{ab}^2}\bm{S}_a\!\cdot \left(2\bm{v}_b-\frac{3}{2}\bm{v}_a\right)\times\bm{n}_{ab},
    \\[2pt]
    V_{\rm SO}^{\rm NLO} 
    &=
    \sum_a\sum_{b\neq a}
    \frac{Gm_b}{r_{ab}^{2}}
    \bigg[
    \bm{S}_a\!\cdot\bm{v}_a\times\bm{n}_{ab}
    \bigg(\!
    -\frac{1}{8}v_a^2
    +3(\bm{v}_a\!\cdot\bm{n}_{ab})
    (\bm{v}_b\!\cdot\bm{n}_{ab})
    -\frac{3}{4}(\bm{v}_b\!\cdot\bm{n}_{ab})^2
    \bigg)
    \notag\\[2pt]
    &\quad\quad
    +\bm{S}_a\!\cdot\bm{v}_b\times\bm{n}_{ab}
    \big(
    v_b^2
    -(\bm{v}_a\!\cdot\bm{v}_b)
    -3(\bm{v}_a\!\cdot\bm{n}_{ab})
    (\bm{v}_b\!\cdot\bm{n}_{ab})
    \big)
    +\bm{S}_a\!\cdot\bm{v}_a\times\bm{v}_b
    \bigg(
    2(\bm{v}_a\!\cdot\bm{n}_{ab})
    -\frac{5}{2}(\bm{v}_b\!\cdot\bm{n}_{ab})
    \bigg)
    \bigg]
    \notag\\[2pt]
    &\quad
    +\sum_a\sum_{b\neq a}
    \frac{G^2m_b}{2r_{ab}^3}
    \Big[
    (m_b-7m_a)\,
    \bm{S}_a\!\cdot\bm{v}_a\times\bm{n}_{ab}
    -7m_b\,
    \bm{S}_a\!\cdot\bm{v}_b\times\bm{n}_{ab}
    \Big]
    \notag\\[2pt]
    &\quad
    +\sum_a\sum_{b\neq a}\sum_{c\neq a,b}
    G^2m_bm_c
    \Bigg\{
    \frac{1}{r_{ab}^2}\,
    \bm{S}_a\!\cdot
    \bigg(
    \frac{3\bm{v}_a+4\bm{v}_b-8\bm{v}_c}{2r_{bc}}
    +\frac{4(\bm{v}_c-\bm{v}_b)}{r_{ac}}
    \bigg)\times\bm{n}_{ab}
    \notag\\[2pt]
    &\quad\quad
    +\frac{1}{r_{ab}r_{bc}^{2}}
    \bigg[\!
    -\bm{S}_a\!\cdot\bm{n}_{ab}\times\bm{n}_{bc}\,
    (\bm{v}_a\!\cdot\bm{n}_{ab})
    +\frac{11}{4}\bm{S}_a\!\cdot\bm{v}_a\times\bm{n}_{bc}
    +\frac{1}{4}
    (\bm{n}_{ab}\!\cdot\bm{n}_{bc})\,
    \bm{S}_a\!\cdot\bm{v}_a\times\bm{n}_{ab}
    \bigg]
    \notag\\[2pt]
    &\quad\quad
    -\frac{1}{r_{ab}r_{ac}^{2}}
    \Big[
    \bm{S}_a\!\cdot\bm{n}_{ab}\times\bm{n}_{ac}\,
    (\bm{v}_b\!\cdot\bm{n}_{ab})
    +3\bm{S}_a\!\cdot\bm{v}_b\times\bm{n}_{ac}
    \Big]
    \notag\\[2pt]
    &\quad\quad
    +\frac{1}{r_{ab}s_{abc}^{2}}
    \Big[
    \bm{S}_a\!\cdot
    (8\bm{v}_c-3\bm{v}_a+\bm{v}_b)\!
    \times(\bm{n}_{ac}+\bm{n}_{bc})
    +\big(
    3(\bm{v}_a\!\cdot\bm{n}_{ab})
    -(\bm{v}_b\!\cdot\bm{n}_{ab})
    -8(\bm{v}_c\!\cdot\bm{n}_{ab})
    \big)\,
    \bm{S}_a\!\cdot\bm{n}_{ab}\times
    (\bm{n}_{ac}+\bm{n}_{bc})
    \Big]
    \notag\\[2pt]
    &\quad\quad
    +\frac{1}{r_{ac}s_{abc}^{2}}
    \Big[
    \bm{S}_a\!\cdot
    (\bm{v}_a-7\bm{v}_c)
    \times(\bm{n}_{ab}-\bm{n}_{bc})
    +\big(
    (\bm{v}_a\!\cdot\bm{n}_{ac})
    -7(\bm{v}_c\!\cdot\bm{n}_{ac})
    \big)\,
    \bm{S}_a\!\cdot
    (\bm{n}_{ab}-\bm{n}_{bc})\times\bm{n}_{ac}
    \Big]
    \notag\\[2pt]
    &\quad\quad
    +\frac{2}{r_{bc}s_{abc}^{2}}
    \Big[\!
    -2\bm{S}_a\!\cdot\bm{v}_a\times\bm{n}_{ab}
    +\bm{S}_a\!\cdot\bm{v}_b\times
    (\bm{n}_{ab}+\bm{n}_{ac})
    +\big(
    (\bm{v}_b\!\cdot\bm{n}_{bc})
    +(\bm{v}_c\!\cdot\bm{n}_{bc})
    -2(\bm{v}_a\!\cdot\bm{n}_{bc})
    \big)\,
    \bm{S}_a\!\cdot\bm{n}_{ab}\times\bm{n}_{bc}
    \Big]
    \notag\\[2pt]
    &\quad\quad
    +\frac{2}{s_{abc}^{3}}
    \Big[
    (\bm{v}_a\!\cdot\bm{n}_{ab})
    -(\bm{v}_a\!\cdot\bm{n}_{ac})
    -2(\bm{v}_a\!\cdot\bm{n}_{bc})
    +3(\bm{v}_b\!\cdot\bm{n}_{ab})
    +4(\bm{v}_b\!\cdot\bm{n}_{ac})
    +(\bm{v}_b\!\cdot\bm{n}_{bc})
    \notag\\[2pt]
    &\quad\quad\quad
    -4(\bm{v}_c\!\cdot\bm{n}_{ab})
    -3(\bm{v}_c\!\cdot\bm{n}_{ac})
    +(\bm{v}_c\!\cdot\bm{n}_{bc})
    \Big]
    \left(
    \bm{S}_a\!\cdot\bm{n}_{ab}\times\bm{n}_{ac}
    +2\bm{S}_a\!\cdot\bm{n}_{ab}\times\bm{n}_{bc}
    \right)
    \Bigg\}.
\end{align}  
\end{widetext}

\section{NLO spin-orbit Hamiltonian}
\label{sec:IV}

In this section we compute the NLO spin-orbit Hamiltonian in the generalized canonical gauge using the LO and NLO spin-orbit potentials calculated in the previous section.

\subsection{Canonical Hamiltonian}

The momentum conjugate to $\bm{x}_a$ is defined by
\begin{equation}
    \bm{p}_a=\frac{\partial L}{\partial \bm{v}_a}.
\end{equation}
To the required order, it is sufficient to use $L=L_{\text{N}}+L_{\text{EIH}}-V_{\text{SO}}^{\text{LO}}$ in this relation. This gives
\begin{align}
\bm{p}_a
&=
m_a\bm{v}_a
+\frac{m_a v_a^2}{2}\bm{v}_a
\notag\\
&\quad
+\sum_{b\neq a}\frac{Gm_a m_b}{r_{ab}}
\left(
3\bm{v}_a
-\frac{7}{2}\bm{v}_b
-\frac{\bm{v}_b\cdot\bm{n}_{ab}}{2}\bm{n}_{ab}
\right)
\notag\\
&\quad
-\sum_{b\neq a}\frac{G}{r_{ab}^2}
\left(
\frac{3}{2}m_b\bm{S}_a\times\bm{n}_{ab}
+2m_a\bm{S}_b\times\bm{n}_{ab}
\right).
\end{align}
Inverting this relation perturbatively, we obtain, to the required order,
\begin{align}
\bm{v}_a
&=
\frac{\bm{p}_a}{m_a}
-\frac{p_a^2}{2m_a^3}\bm{p}_a
\notag\\
&\quad
+\sum_{b\neq a}\frac{G}{r_{ab}}
\left(
-3\frac{m_b}{m_a}\bm{p}_a
+\frac{7}{2}\bm{p}_b
+\frac{\bm{p}_b\!\cdot\bm{n}_{ab}}{2}\bm{n}_{ab}
\right)
\notag\\
&\quad
+\sum_{b\neq a}
\frac{G}{m_a r_{ab}^2}
\left(
\frac{3}{2}m_b\bm{S}_a\times\bm{n}_{ab}
+2m_a\bm{S}_b\times\bm{n}_{ab}
\right).
\end{align}
The Hamiltonian is obtained by the Legendre transform
\begin{equation}
    H= \sum_{a=1}^N \bm{p}_{a}\!\cdot\bm{v}_a-L .
\end{equation}
We evaluate this expression with $L=L_{\rm N}+L_{\rm EIH}-V_{\rm SO}^{\rm LO}-V^{\rm NLO}_{\rm SO}$,  and substitute the velocities in terms of the momenta. The terms linear in spin give the NLO spin-orbit Hamiltonian,
\begin{widetext}
\begin{align}
H_{\rm SO}^{\rm NLO}
&=
\sum_a\sum_{b\neq a}
\frac{G}{r_{ab}^2}
\bigg[
\left(\!
-\frac{5m_b\,\bm{p}_a^2}{8m_a^3}
-\frac{3(\bm{n}_{ab}\!\cdot\bm{p}_a)(\bm{n}_{ab}\!\cdot\bm{p}_b)}{m_a^2}
+\frac{3(\bm{n}_{ab}\!\cdot\bm{p}_b)^2}{4m_am_b}
\right)
\bm{S}_a\!\cdot\bm{n}_{ab}\times\bm{p}_a
\notag\\[2pt]
&\quad\quad
+\left(
\frac{\bm{p}_a\!\cdot\bm{p}_b}{m_am_b}
+\frac{3(\bm{n}_{ab}\!\cdot\bm{p}_a)(\bm{n}_{ab}\!\cdot\bm{p}_b)}{m_am_b}
\right)
\bm{S}_a\!\cdot\bm{n}_{ab}\times\bm{p}_b
+\left(
\frac{2(\bm{n}_{ab}\!\cdot\bm{p}_a)}{m_a^2}
-\frac{5(\bm{n}_{ab}\!\cdot\bm{p}_b)}{2m_am_b}
\right)
\bm{S}_a\!\cdot\bm{p}_a\times\bm{p}_b
\bigg]
\notag\\[2pt]
&\quad
+\sum_a\sum_{b\neq a}
\frac{G^2}{r_{ab}^3}
\bigg[\!
-\left(
\frac{7}{2}m_b+\frac{5m_b^2}{m_a}
\right)
\bm{S}_a\!\cdot\bm{n}_{ab}\times\bm{p}_a
+\left(
6m_a+\frac{35}{4}m_b
\right)
\bm{S}_a\!\cdot\bm{n}_{ab}\times\bm{p}_b
\bigg]
\notag\\[2pt]
&\quad
+\sum_a\sum_{b\neq a}\sum_{c\neq a,b}
G^2
\Bigg\{\!
-\frac{3m_bm_c}{2m_ar_{ac}^2}
\left(
\frac{3}{r_{ab}}+\frac{1}{r_{bc}}
\right)
\bm{S}_a\!\cdot\bm{n}_{ac}\times\bm{p}_a
+\frac{m_c}{r_{ac}^2}
\left(
\frac{17}{4r_{ab}}-\frac{3}{r_{bc}}
\right)
\bm{S}_a\!\cdot\bm{n}_{ac}\times\bm{p}_b
\notag\\[2pt]
&\quad\quad
+\frac{4m_b}{r_{ac}^2}
\left(
\frac{1}{r_{ab}}+\frac{1}{r_{bc}}
\right)
\bm{S}_a\!\cdot\bm{n}_{ac}\times\bm{p}_c
-\frac{11m_bm_c}{4m_ar_{ab}r_{bc}^2}
\bm{S}_a\!\cdot\bm{n}_{bc}\times\bm{p}_a
+\frac{7m_b}{4r_{ab}^2r_{ac}}
(\bm{n}_{ac}\!\cdot\bm{p}_c)\,
\bm{S}_a\!\cdot\bm{n}_{ab}\times\bm{n}_{ac}
\notag\\[2pt]
&\quad\quad
-\bigg(
\frac{m_b}{r_{ab}^2r_{bc}}
(\bm{n}_{bc}\!\cdot\bm{p}_c)
+\frac{m_bm_c}{m_ar_{ab}r_{bc}^2}
(\bm{n}_{ab}\!\cdot\bm{p}_a)
\bigg)
\bm{S}_a\!\cdot\bm{n}_{ab}\times\bm{n}_{bc}
-\frac{m_bm_c}{4m_ar_{ab}r_{bc}^2}
(\bm{n}_{ab}\!\cdot\bm{n}_{bc})\,
\bm{S}_a\!\cdot\bm{n}_{ab}\times\bm{p}_a
\notag\\[2pt]
&\quad\quad
+\frac{1}{s_{abc}^2}
\bigg[
\frac{2m_bm_c}{m_ar_{ab}}\,
\bm{S}_a\!\cdot(\bm{n}_{ac}+\bm{n}_{bc})\times\bm{p}_a
+\frac{6m_c}{r_{ab}}\,
\bm{S}_a\!\cdot(\bm{n}_{ac}+\bm{n}_{bc})\times\bm{p}_b
\notag\\[2pt]
&\quad\quad\quad
-\frac{8m_b}{r_{ab}}\,
\bm{S}_a\!\cdot(\bm{n}_{ac}+\bm{n}_{bc})\times\bm{p}_c
+\frac{4m_bm_c}{m_ar_{bc}}\,
\bm{S}_a\!\cdot\bm{n}_{ab}\times\bm{p}_a
-2\,\frac{m_c}{r_{bc}}\,
\bm{S}_a\!\cdot(\bm{n}_{ab}+\bm{n}_{ac})\times\bm{p}_b
\notag\\[2pt]
&\quad\quad\quad
+\bm{S}_a\!\cdot\bm{n}_{ab}\times\bm{n}_{ac}
\bigg(
\frac{2m_bm_c}{m_ar_{ab}}
(\bm{n}_{ab}\!\cdot\bm{p}_a)
+\frac{6m_c}{r_{ab}}
(\bm{n}_{ab}\!\cdot\bm{p}_b)
+\frac{8m_c}{r_{ac}}
(\bm{n}_{ac}\!\cdot\bm{p}_b)
\bigg)
\notag\\[2pt]
&\quad\quad\quad
+2\bm{S}_a\!\cdot\bm{n}_{ab}\times\bm{n}_{bc}
\bigg(
\frac{m_bm_c}{m_ar_{ab}}
(\bm{n}_{ab}\!\cdot\bm{p}_a)
-\frac{2m_bm_c}{m_ar_{bc}}
(\bm{n}_{bc}\!\cdot\bm{p}_a)
\notag\\[2pt]
&\quad\quad\quad\quad
+\frac{3m_c}{r_{ab}}
(\bm{n}_{ab}\!\cdot\bm{p}_b)
+\frac{m_c}{r_{bc}}
(\bm{n}_{bc}\!\cdot\bm{p}_b)
-\frac{4m_b}{r_{ab}}
(\bm{n}_{ab}\!\cdot\bm{p}_c)
+\frac{m_b}{r_{bc}}
(\bm{n}_{bc}\!\cdot\bm{p}_c)
\bigg)
\bigg]
\notag\\[2pt]
&\quad\quad
+\frac{1}{s_{abc}^3}
\bigg[
\bm{S}_a\!\cdot\bm{n}_{ab}\times\bm{n}_{ac}
\bigg(
\frac{4m_bm_c}{m_a}
(\bm{n}_{ab}\!\cdot\bm{p}_a)
-\frac{4m_bm_c}{m_a}
(\bm{n}_{bc}\!\cdot\bm{p}_a)
\notag\\[2pt]
&\quad\quad\quad\quad
+12m_c(\bm{n}_{ab}\!\cdot\bm{p}_b)
+16m_c(\bm{n}_{ac}\!\cdot\bm{p}_b)
+4m_c(\bm{n}_{bc}\!\cdot\bm{p}_b)
\bigg)
\notag\\[2pt]
&\quad\quad\quad
+2\bm{S}_a\!\cdot\bm{n}_{ab}\times\bm{n}_{bc}
\bigg(
\frac{m_bm_c}{m_a}
\left(
2\,\bm{n}_{ab}\!\cdot\bm{p}_a
-2\,\bm{n}_{ac}\!\cdot\bm{p}_a
-4\,\bm{n}_{bc}\!\cdot\bm{p}_a
\right)
\notag\\[2pt]
&\quad\quad\quad\quad
+m_c
\left(
6\,\bm{n}_{ab}\!\cdot\bm{p}_b
+8\,\bm{n}_{ac}\!\cdot\bm{p}_b
+2\,\bm{n}_{bc}\!\cdot\bm{p}_b
\right)
-m_b
\left(
8\,\bm{n}_{ab}\!\cdot\bm{p}_c
+6\,\bm{n}_{ac}\!\cdot\bm{p}_c
-2\,\bm{n}_{bc}\!\cdot\bm{p}_c
\right)
\bigg)
\bigg]
\Bigg\}.
\end{align}
\end{widetext}

\subsection{Comparison with ADM Hamiltonian}

We compare our result with the $N$-body spin-orbit Hamiltonian of Hartung and Steinhoff~\cite{hartung2011}, obtained in the ADM formalism. We follow the same procedure used in~\cite{levi2010so} to compare the binary EFT Hamiltonian with the canonical ADM Hamiltonian of~\cite{damour2008}. We define
\begin{equation}
    \Delta H = H_{\text{EFT}}-H_{\text{HS}},
\end{equation}
where $H_{\rm EFT}$ is our NLO spin-orbit Hamiltonian and $H_{\rm HS}$ is the Hamiltonian of Hartung and Steinhoff~\cite{hartung2011}. In our notation, the latter is written as
\begin{widetext}
\begin{align}
H_{\text{HS}}
&=
\sum_a \sum_{b\neq a}\frac{G}{r_{ab}^2}
\bigg[
\left(
\frac{3\,\bm{p}_b^2}{4m_am_b}
-\frac{3\,(\bm{n}_{ab}\!\cdot\bm{p}_b)^2}{2m_am_b}
-\frac{5m_b\,\bm{p}_a^2}{8m_a^3}
-\frac{3\,(\bm{p}_a\!\cdot\bm{p}_b)}{4m_a^2}
-\frac{3\,(\bm{n}_{ab}\!\cdot\bm{p}_a)(\bm{n}_{ab}\!\cdot\bm{p}_b)}{4m_a^2}
\right)
\bm{S}_a\!\cdot\bm{n}_{ab}\times\bm{p}_a
\notag\\[2pt]
&\quad\quad
+\left(
\frac{\bm{p}_a\!\cdot\bm{p}_b}{m_am_b}
+\frac{3\,(\bm{n}_{ab}\!\cdot\bm{p}_a)(\bm{n}_{ab}\!\cdot\bm{p}_b)}{m_am_b}
\right)
\bm{S}_a\!\cdot\bm{n}_{ab}\times\bm{p}_b
+\left(
\frac{3\,(\bm{n}_{ab}\!\cdot\bm{p}_a)}{4m_a^2}
-\frac{2\,(\bm{n}_{ab}\!\cdot\bm{p}_b)}{m_am_b}
\right)
\bm{S}_a\!\cdot\bm{p}_a\times\bm{p}_b
\bigg]
\notag\\[2pt]
&\quad
+\sum_a \sum_{b\neq a}\frac{G^2}{r_{ab}^3}
\bigg[\!
-\left(
\frac{11}{2}m_b+\frac{5m_b^2}{m_a}
\right)
\bm{S}_a\!\cdot\bm{n}_{ab}\times\bm{p}_a
+\left(
6m_a+\frac{15}{2}m_b
\right)
\bm{S}_a\!\cdot\bm{n}_{ab}\times\bm{p}_b
\bigg]\notag\\[2pt]
&\quad
-\sum_a\sum_{b\neq a}\sum_{c\neq a,b}
\frac{G^2}{r_{ab}^2}
\left(
\frac{5}{r_{ac}}+\frac{1}{r_{bc}}
\right)
\frac{m_bm_c}{m_a}\,
\bm{S}_a\!\cdot\bm{n}_{ab}\times\bm{p}_a
\notag\\[2pt]
&\quad
+\sum_a\sum_{b\neq a}\sum_{c\neq a,b}
G^2m_a
\Bigg\{
\frac{1}{s_{abc}}
\bigg[
\left(
-\frac{8}{r_{bc}^2}
+\frac{8}{r_{ab}r_{ac}}
-\frac{4}{r_{ab}r_{bc}}
-\frac{4}{r_{ac}r_{bc}}
-\frac{4r_{ab}}{r_{ac}r_{bc}^2}
-\frac{4r_{ac}}{r_{ab}r_{bc}^2}
\right)
\bm{S}_c\!\cdot\bm{n}_{bc}\times\bm{p}_b
\notag\\[2pt]
&\quad\quad\quad
+\left(
\frac{3}{r_{ab}r_{ac}}
-\frac{6}{r_{ab}r_{bc}}
-\frac{3}{r_{ac}r_{bc}}
-\frac{3r_{ab}}{r_{ac}^2r_{bc}}
+\frac{3r_{bc}}{r_{ab}r_{ac}^2}
\right)
\bm{S}_c\!\cdot\bm{n}_{ac}\times\bm{p}_b
\bigg]
\notag\\[2pt]
&\quad\quad
+\frac{1}{s_{abc}^2}\,
\bm{S}_c\!\cdot\bm{n}_{ac}\times\bm{n}_{bc}
\bigg[
\frac{16\,(\bm{n}_{ac}\!\cdot\bm{p}_b)}{r_{ab}}
\notag\\[2pt]
&\quad\quad\quad
+\left(
\frac{4}{r_{ab}}
-\frac{2r_{bc}}{r_{ab}r_{ac}}
+\frac{2r_{ac}}{r_{ab}^2}
-\frac{2r_{bc}^2}{r_{ab}^2r_{ac}}
+\frac{8r_{ab}}{r_{ac}^2}
+\frac{7r_{bc}}{r_{ac}^2}
-\frac{2r_{bc}^2}{r_{ab}r_{ac}^2}
+\frac{r_{bc}}{r_{ab}^2}
-\frac{r_{bc}^3}{r_{ab}^2r_{ac}^2}
\right)
(\bm{n}_{ab}\!\cdot\bm{p}_b)
\notag\\[2pt]
&\quad\quad\quad
+\left(
\frac{12}{r_{ab}}
-\frac{2}{r_{ac}}
+\frac{5}{r_{bc}}
-\frac{2r_{ab}}{r_{ac}^2}
+\frac{7r_{bc}}{r_{ac}^2}
-\frac{2r_{ab}}{r_{ac}r_{bc}}
+\frac{6r_{ac}}{r_{ab}r_{bc}}
-\frac{r_{ab}^2}{r_{ac}^2r_{bc}}
+\frac{8r_{bc}^2}{r_{ab}r_{ac}^2}
\right)
(\bm{n}_{bc}\!\cdot\bm{p}_b)
\bigg]
\Bigg\}
\notag\\[2pt]
&\quad
+\sum_a\sum_{b\neq a}\sum_{c\neq a,b}
G^2\frac{m_am_b}{m_c}
\Bigg\{
\frac{1}{s_{abc}^2}
\bigg[\!
-\frac{1}{r_{ac}}
-\frac{1}{r_{ab}}
\left(
2+\frac{r_{bc}}{2r_{ac}}
\right)
+\frac{1}{r_{ab}^2}
\left(
-4r_{ac}+r_{bc}+\frac{r_{bc}^2}{r_{ac}}
\right)
\notag\\[2pt]
&\quad\quad\quad
+\frac{1}{r_{ab}^3}
\left(\!
-2r_{ac}^2+r_{bc}^2-\frac{3}{2}r_{ac}r_{bc}
+\frac{r_{bc}^3}{2r_{ac}}
\right)
\bigg]
(\bm{n}_{ac}\!\cdot\bm{p}_c)\,
\bm{S}_c\!\cdot\bm{n}_{ac}\times\bm{n}_{bc}
\notag\\[2pt]
&\quad\quad
+\frac{1}{s_{abc}}
\bigg[
\frac{1}{8r_{ac}^2}
-\frac{1}{4r_{ac}r_{bc}}
-\frac{r_{ab}}{4r_{ac}^2r_{bc}}
+\frac{1}{r_{ab}}
\left(
\frac{3}{8r_{ac}}-\frac{1}{r_{bc}}+\frac{3r_{bc}}{8r_{ac}^2}
\right)
\notag\\[2pt]
&\quad\quad\quad
+\frac{1}{r_{ab}^2}
\left(\!
-\frac{5}{8}+\frac{3r_{ac}}{4r_{bc}}-\frac{r_{bc}^2}{8r_{ac}^2}
\right)
+\frac{1}{r_{ab}^3}
\left(
\frac{r_{ac}}{8}-\frac{5r_{bc}}{8}+\frac{3r_{ac}^2}{4r_{bc}}
-\frac{r_{bc}^2}{8r_{ac}}-\frac{r_{bc}^3}{8r_{ac}^2}
\right)
\bigg]
\bm{S}_c\!\cdot\bm{n}_{ac}\times\bm{p}_c
\notag\\[2pt]
&\quad\quad
+(a\leftrightarrow b)
\Bigg\},
\end{align}
\end{widetext}
where $(a\leftrightarrow b)$ denotes the contribution obtained by exchanging the object labels $a$ and $b$ in all preceding terms.
The difference between our Hamiltonian and the ADM result is
\begin{widetext}
\begin{align}
\Delta H
&=
\sum_a \sum_{b\neq a}\frac{Gm_b}{r_{ab}^{2}}\,
\bm{S}_a\!\cdot\!
\Bigg[
\frac{\bm{p}_a\times\bm{n}_{ab}}{m_a}
\Bigg(\!
-\frac{3\,\bm{p}_a\!\cdot\bm{p}_b}{4m_am_b}
+\frac{3\,p_b^2}{4m_b^2}
+\frac{9\,(\bm{p}_a\!\cdot\bm{n}_{ab})(\bm{p}_b\!\cdot\bm{n}_{ab})}{4m_am_b}
-\frac{9\,(\bm{p}_b\!\cdot\bm{n}_{ab})^2}{4m_b^2}
-\frac{2Gm_a}{r_{ab}}
\Bigg)
\notag\\[2pt]
&\quad\quad
+\frac{\bm{p}_a\times\bm{p}_b}{m_am_b}
\Bigg(
\frac{5\,\bm{p}_a\!\cdot\bm{n}_{ab}}{4m_a}
-\frac{\bm{p}_b\!\cdot\bm{n}_{ab}}{2m_b}
\Bigg)
\Bigg]
\notag\\[2pt]
&\quad-\sum_a \sum_{b\neq a}\frac{5G^2m_b}{4r_{ab}^3}
\bm{S}_a\!\cdot\bm{p}_b\times\bm{n}_{ab}
,\notag\\[2pt]
&\quad
+\sum_a \sum_{b\neq a}\sum_{c\neq a,b}
G^2\bm{S}_a\!\cdot\!
\Bigg[
\frac{\bm{p}_a\times\bm{n}_{ab}}{m_a}
\Bigg(
\frac{5m_bm_c\,r_{ab}}{4r_{ac}r_{bc}^3}
-\frac{13m_bm_c}{8r_{bc}^3}
-\frac{3m_bm_c}{8r_{ab}^2r_{bc}}
\left(
1-\frac{r_{ac}^2}{r_{bc}^2}
\right)
\Bigg)
\notag\\[2pt]
&\quad\quad
+\frac{3m_b}{4r_{ab}^2r_{ac}}\,
(\bm{n}_{ac}\!\cdot\bm{p}_c)\,
\bm{n}_{ab}\times\bm{n}_{ac}
-\frac{5m_c}{4r_{ab}r_{ac}^2}\,\bm{p}_b\times\bm{n}_{ac}
\Bigg].
\end{align}
\end{widetext}

\noindent
Throughout our calculations, we used the relation
\begin{align}
    \bm{n}_{bc}=\frac{r_{ac}}{r_{bc}}\bm{n}_{ac}-\frac{r_{ab}}{r_{bc}}\bm{n}_{ab},
\end{align}
to eliminate one of the unit vectors. 
Scalar products of unit vectors were removed using $\bm{n}_{ab}\!\cdot\bm{n}_{ab}=1$ and
\begin{align}
    \bm{n}_{ac}\!\cdot\bm{n}_{bc}
    &= \frac{r_{ac}^{2}+r_{bc}^{2}-r_{ab}^{2}}{2r_{ac}r_{bc}}, \\[2pt]
    \bm{n}_{ab}\!\cdot\bm{n}_{bc}
    &= -\frac{r_{ab}^{2}+r_{bc}^{2}-r_{ac}^{2}}{2r_{ab}r_{bc}}, \\[2pt]
    \bm{n}_{ab}\!\cdot\bm{n}_{ac}
    &= \frac{r_{ab}^{2}+r_{ac}^{2}-r_{bc}^{2}}{2r_{ab}r_{ac}}.
\end{align}

For the EFT Hamiltonian to produce the same equations of motion as the ADM Hamiltonian of Hartung and Steinhoff, there must exist a generator of canonical transformation $g$ satisfying 
\begin{equation}
    \Delta H= \left\{H,g\right\} = -\frac{dg}{dt}.
    \label{leep}
\end{equation}
Since $\Delta H\sim \mathcal{O}(GSp^3)+\mathcal{O}(G^2Sp)$, the generator must be of order $\mathcal{O}(GSp^2)$. Differentiating its coordinate dependence gives the $\mathcal{O}(GSp^3)$ terms, while differentiating the momenta gives the $\mathcal{O}(G^2Sp)$ terms. We therefore use the following general generator~\cite{levi2010so}
\begin{widetext}
\begin{align}
g
&=
\sum_a \sum_{b\neq a}\frac{Gm_b}{r_{ab}}\,
\bm{S}_a \!\cdot
\Bigg[
g_1\,\frac{\bm{p}_a \times \bm{p}_b}{m_a m_b}
+\frac{\bm{p}_a \times \bm{n}_{ab}}{m_a}
\left(
g_2\,\frac{\bm{p}_a \!\cdot \bm{n}_{ab}}{m_a}
+g_3\,\frac{\bm{p}_b \!\cdot \bm{n}_{ab}}{m_b}
\right)
\notag\\[2pt]
&\qquad\qquad
+\frac{\bm{p}_b \times \bm{n}_{ab}}{m_b}
\left(
g_4\,\frac{\bm{p}_a \!\cdot \bm{n}_{ab}}{m_a}
+g_5\,\frac{\bm{p}_b \!\cdot \bm{n}_{ab}}{m_b}
\right)
\Bigg].
\label{generator}
\end{align}
\end{widetext}
To the desired order, we can compute $dg/dt$ using the time derivatives
\begin{align}
\dot{\bm{p}}_a &= -\sum_{b\neq a}\frac{Gm_am_b}{r_{ab}^2}\bm{n}_{ab}, \\[2pt]
\dot{r}_{ab} &= \frac{\bm{p}_a\!\cdot\bm{n}_{ab}}{m_a}-\frac{\bm{p}_b\!\cdot\bm{n}_{ab}}{m_b}, \\[2pt]
\dot{\bm{n}}_{ab} &= \frac{1}{r_{ab}}
\left(
\frac{\bm{p}_a}{m_a}-\frac{\bm{p}_b}{m_b}-\bm{n}_{ab}\,\dot{r}_{ab}
\right), \\[2pt]
\dot{S}_a^{ij} &= 0.
\end{align}
Substituting the above formulas into Eq.~\eqref{leep}, we find
\begin{equation}
    g_1=5/4, \quad g_3=3/4,\quad g_2=g_4=g_5=0.
\end{equation}
This is the same solution found in the binary case in~\cite{levi2015spinning}. In this way, we have shown that our Hamiltonian is physically equivalent to the ADM Hamiltonian of~\cite{hartung2011}.
\section{Covariant SSC derivation}
\label{sec:V}

In this section we repeat the analysis of Secs.~\ref{sec:III} and~\ref{sec:IV} using the covariant SSC instead of the generalized canonical gauge. We first give the new Feynman rules. We then compute the diagrams contributing to the LO and NLO spin-orbit sectors, derive the NLO spin-orbit potential, obtain the canonical Hamiltonian, and compare it with the ADM Hamiltonian of~\cite{hartung2011}.

\subsection{Feynman rules}

In the covariant-SSC approach, the only rules that differ from those in Table~\ref{tab:Feynman} are the worldline spin vertices. These are listed in Table~\ref{tab:Feynman_old}. Note that there is one additional two-graviton rule, which was not considered in Table~\ref{tab:Feynman} because it contributes only at higher orders in the generalized canonical gauge.
\begin{table*}[t]
\caption{Worldline spin vertices for the redundant spin variables. Compared with the generalized canonical gauge, there is one additional vertex, which contributes only at higher order in the generalized canonical gauge.}
\label{tab:Feynman_old}
\centering

\begin{tabular}{l}
\hline\hline
\noalign{\vskip 2pt}
\multicolumn{1}{c}{Worldline spin vertices} \\
\noalign{\vskip 1pt}
\hline
\noalign{\vskip 8pt}

\multicolumn{1}{@{}c@{}}{%
\parbox{0.75\textwidth}{\centering
$\displaystyle
\begin{aligned}
&\vcenter{\hbox{
\begin{tikzpicture}[
    baseline=(current bounding box.center),
    x=0.8cm,
    y=0.8cm,
    line cap=round,
    line join=round
]
    \draw[worldline] (0,-0.8) -- (0,0.8);
    \draw[phi] (0,0) -- (1,0);
    \draw[black, line width=0.6pt, fill=black] (0,0) circle (2.5pt);
\end{tikzpicture}}}
&&\!\!\!=
\int dt\,\Big(
S^{ij}v^i\partial_j \phi
+S^{0i}\partial_i\phi
-S^{0i}v^i\partial_0\phi
\Big)
\\[4pt]
&\vcenter{\hbox{
\begin{tikzpicture}[
    baseline=(current bounding box.center),
    x=0.8cm,
    y=0.8cm,
    line cap=round,
    line join=round
]
    \draw[worldline] (0,-0.8) -- (0,0.8);
    \draw[A] (0,0) -- (1,0);
    \draw[black, line width=0.6pt, fill=black] (0,0) circle (2.5pt);
\end{tikzpicture}}}
&&\!\!\!=
\int dt\,\bigg(
\frac{1}{2}S^{ij}\partial_i A_j
-\frac{1}{2}S^{0i}v^j\partial_i A_j
+\frac{1}{2}S^{0i}\partial_0 A_i
\bigg)
\\[4pt]
&\vcenter{\hbox{
\begin{tikzpicture}[
    baseline=(current bounding box.center),
    x=0.8cm,
    y=0.8cm,
    line cap=round,
    line join=round
]
    \draw[worldline] (0,-0.8) -- (0,0.8);
    \draw[sigma] (0,0) -- (1,0);
    \draw[black, line width=0.6pt, fill=black] (0,0) circle (2.5pt);
\end{tikzpicture}}}
&&\!\!\!=
\int dt\,
\frac{1}{2}S^{ij}v^k \partial_i \sigma_{jk}
\\[4pt]
&\vcenter{\hbox{
\begin{tikzpicture}[
    baseline=(current bounding box.center),
    x=0.8cm,
    y=0.8cm,
    line cap=round,
    line join=round
]
    \draw[worldline] (0,-0.8) -- (0,0.8);
    \draw[phi] (0,0) -- ({cos(30)},{sin(30)});
    \draw[phi] (0,0) -- ({cos(30)},{-sin(30)});
    \draw[black, line width=0.6pt, fill=black] (0,0) circle (2.5pt);
\end{tikzpicture}}}
&&\!\!\!=
\int dt\,
2S^{0i}\phi\,\partial_i\phi
\\[4pt]
&\vcenter{\hbox{
\begin{tikzpicture}[
    baseline=(current bounding box.center),
    x=0.8cm,
    y=0.8cm,
    line cap=round,
    line join=round
]
    \draw[worldline] (0,-0.8) -- (0,0.8);
    \draw[A] (0,0) -- ({cos(30)},{sin(30)});
    \draw[phi] (0,0) -- ({cos(30)},{-sin(30)});
    \draw[black, line width=0.6pt, fill=black] (0,0) circle (2.5pt);
\end{tikzpicture}}}
&&\!\!\!=
\int dt\,
\Big(
2S^{ij}\phi\,\partial_i A_j
-\frac{1}{2}S^{ij}A_i\partial_j \phi
\Big)
\end{aligned}
$
}}\\[6pt]

\noalign{\vskip 12pt}
\hline\hline
\end{tabular}

\end{table*}

\subsection{Diagrams}

\begin{figure}[H]
\centering

\begin{tabular}{@{}c@{\hspace{8mm}}c@{}}

\begin{tikzpicture}[x=1cm,y=1cm, line cap=round, scale=0.8, transform shape]
  \draw[phi] (1,0) -- (2,2);
  \draw[phi] (3,0) -- (2,2);
  \draw[worldline] (0.5,0) -- (3.5,0);
  \draw[worldline] (0.5,2) -- (3.5,2);

  \draw[black, line width=0.6pt, fill=white] (1,0) circle (2.5pt);
  \draw[black, line width=0.6pt, fill=white] (3,0) circle (2.5pt);
  \draw[black, line width=0.6pt, fill=black] (2,2) circle (2.5pt);

  \node[above] at (2,2) {$v^3$};
  \node[below] at (1,0) {$v^0$};
  \node[below] at (3,0) {$v^0$};
\end{tikzpicture}
&
\begin{tikzpicture}[x=1cm,y=1cm, line cap=round, scale=0.75, transform shape]
  % interaction
  \draw[phi] (-1.2,-0.6928) -- (0,1.3856);
  \draw[phi] (1.2,-0.6928) -- (0,1.3856);

  % black lines
  \draw[worldline] (-0.8,1.3856) -- (0.8,1.3856);
  \draw[worldline] (-1.6,0) -- (-0.8,-1.3856);
  \draw[worldline] (0.8,-1.3856) -- (1.6,0);

  % worldline vertices
  \draw[black, line width=0.6pt, fill=white] (-1.2,-0.6928) circle (2.5pt);
  \draw[black, line width=0.6pt, fill=white] (1.2,-0.6928) circle (2.5pt);

  % spin coupling
  \draw[black, line width=0.6pt, fill=black] (0,1.3856) circle (2.5pt);

  % labels
  \node[above] at (0,1.3856) {$v^3$};
  \node[below] at (-1.5,-0.55) {$v^0$};
  \node[below] at (1.5,-0.55) {$v^0$};
\end{tikzpicture}
\\[-1mm]
(n1) & (n2)

\end{tabular}

\caption{Additional two-body and three-body diagrams contributing to the NLO spin-orbit sector in the SSC approach, relative to the generalized canonical gauge diagrams in Figs.~\ref{fig:NLO_SO_diagrams} and~\ref{fig:NLOthree}. The black and white circles denote spin and mass couplings, respectively. Both are two-graviton diagrams with symmetry factor $2$. }
\label{fig:SSC}
\end{figure}
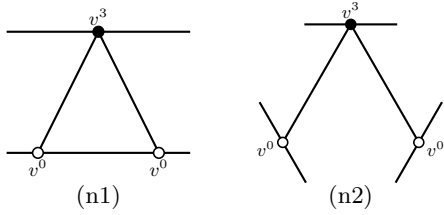

The additional $\phi\phi$ worldline spin vertex gives one new generic worldline diagram. This leads to two extra NLO diagrams, denoted by (n1) and (n2): one two-body diagram and one three-body diagram. They are shown in Fig.~\ref{fig:SSC}.

We include in our calculation the same diagrams as in the generalized canonical-gauge calculation, together with the additional diagrams (n1) and (n2). However, some diagram values change. The modified diagrams are (a1) at LO, and (b1), (b2), (b3), (b5), (c1), (c3), (e1) and (e3) at NLO. The values of these diagrams, together with the new contributions (n1) and (n2), are
\begin{widetext}
\begin{align}
    L^{(\rm a1)} &= 
    \sum_a \sum_{b\neq a}\frac{Gm_b}{r_{ab}^2}
    \big(
    \bm{S}_a\!\cdot \bm{v}_a\times \bm{n}_{ab}+S_a^{0i}n_{ab}^i
    \big),
    \\[2pt]
    L^{(\rm b1)} &=
     \sum_a \sum_{b\neq a}\frac{3 G m_b}{2 r_{ab}^{2}}
    \left( \bm{S}_a \!\cdot \bm{v}_a \times \bm{n}_{ab} + S_a^{0i} n_{ab}^{i} \right)\bm{v}_b^{2},
    \\[2pt]
    L^{(\rm b2)} &=
    \sum_a\sum_{b\neq a}\frac{G m_b}{r_{ab}^{2}} S_a^{0i} n_{ab}^{i},
    \\[2pt]
    L^{(\rm b3)} &=
    \sum_a\sum_{b\neq a}
    \frac{G m_b}{2 r_{ab}^{2}} 
    \Big[
    \bm{S}_a \!\cdot\bm{v}_a \times \bm{n}_{ab}(\bm{v}_a \!\cdot \bm{v}_b)
    + \bm{S}_a \!\cdot\bm{v}_a \times \bm{v}_b(\bm{v}_a \!\cdot \bm{n}_{ab})
    - 3 \bm{S}_a \!\cdot\bm{v}_a \times \bm{n}_{ab}(\bm{v}_a \!\cdot \bm{n}_{ab})(\bm{v}_b \!\cdot \bm{n}_{ab})
    \Big]
    \notag\\
    &\quad
    +\sum_a\sum_{b\neq a} \frac{G m_b}{2 r_{ab}} \Big[ \bm{S}_a \!\cdot \bm{v}_b \times \bm{a}_a
    +\bm{S}_a\!\cdot \bm{a}_a\times \bm{n}_{ab}\,(\bm{v}_b\!\cdot\bm{n}_{ab}) \Big]
    \notag\\
    &\quad
    +\sum_a\sum_{b\neq a} \frac{G m_b}{2 r_{ab}^{2}}
    \Big[
    S_a^{0i} n_{ab}^{i}(\bm{v}_a \!\cdot \bm{v}_b)
    + S_a^{0i} v_b^{i}(\bm{v}_a \!\cdot \bm{n}_{ab})
    - 3 S_a^{0i} n_{ab}^{i}(\bm{v}_a \!\cdot \bm{n}_{ab})(\bm{v}_b \!\cdot \bm{n}_{ab})
    \Big]
    \notag\\
    &\quad
    - \sum_a\sum_{b\neq a}\frac{G m_b}{2 r_{ab}} \partial_t S_a^{0i} v_b^{i},
    \\[2pt]
    L^{(\rm b5)} &=
    -\sum_a\sum_{b\neq a}
    \frac{2 G m_b}{r_{ab}^{2}}
    \big(
    S_a^{0i} n_{ab}^{i}(\bm{v}_a \!\cdot \bm{v}_b)
    + S_a^{0i} v_b^{i}(\bm{v}_a \!\cdot \bm{n}_{ab})
    \big)
    -\sum_a\sum_{b\neq a} \frac{2 G m_b}{r_{ab}} \partial_t S_a^{0i} v_b^{i},
    \\[2pt]
    L^{(\rm c1)} &=
    \sum_a\sum_{b\neq a}\frac{10 G^{2} m_b^{2}}{r_{ab}^{3}} \bm{S}_a \!\cdot \bm{v}_b \times \bm{n}_{ab},
    \\[2pt]
    L^{(\rm c3)} &=
    -\sum_a\sum_{b\neq a}\frac{G^{2} m_a m_b}{r_{ab}^{3}}
    \left[\bm{S}_a \!\cdot \bm{v}_a \times \bm{n}_{ab} + S_a^{0i} n_{ab}^{i}\right],
    \\[2pt]
    L^{(\rm e1)} &=
    \sum_a\sum_{b\neq a}\sum_{c\neq a, b}\frac{2G^2m_b m_c}{r_{ab}r_{ac}}\bm{S}_{a}\!\cdot \bm{v}_b\times\left(4\frac{\bm{n}_{ab}}{r_{ab}}+\frac{\bm{n}_{ac}}{r_{ac}}\right),
    \\[2pt]
    L^{(\rm e3)} &=
    -\sum_a\sum_{b\neq a}\sum_{c\neq a,b} \frac{G^2m_b m_c}{r_{ab}^2 r_{bc}}\big(\bm{S}_a\!\cdot \bm{v}_a\times \bm{n}_{ab}+S_a^{0i}n_{ab}^i\big),
    \\[2pt]
    L^{(\rm n1)} &=
    -\sum_a\sum_{b\neq a}\frac{2 G^{2} m_b^{2}}{r_{ab}^{3}} S_a^{0i} n_{ab}^{i},
    \\[2pt]
    L^{(\rm n2)} &=
    -\sum_a\sum_{b\neq a}\sum_{c\neq a,b} \frac{G^2m_b m_c}{r_{ab}r_{ac}}S_a^{0i}\left(\frac{n_{ab}^i}{r_{ab}}+\frac{n^i_{ac}}{r_{ac}}\right).
\end{align}
\end{widetext}

\subsection{Potential}

Summing all diagrams contributing at LO and NLO, we obtain the potentials
\begin{widetext}
\begin{align}
    V^{\rm LO}_{\rm SO} 
    &= 
    \sum_a \sum_{b\neq a}\frac{Gm_b}{r_{ab}^2}\Big[
    \bm{S}_a\!\cdot (2\bm{v}_b-\bm{v}_a)\times\bm{n}_{ab}
    - S^{0i}_a n^i_{ab}
    \Big], \\[2pt]
    V^{\rm NLO}_{\rm SO} 
    &=
    \sum_a\sum_{b\neq a}
    \frac{Gm_b}{r_{ab}}
    \bigg[
    \bm{S}_a\!\cdot\bm{a}_b\times
    \big(
    \bm{v}_a-\bm{n}_{ab}(\bm{v}_a\!\cdot\bm{n}_{ab})
    \big)
    +\frac{1}{2}\bm{S}_a\!\cdot\bm{a}_a\times
    \big(
    \bm{v}_b-\bm{n}_{ab}(\bm{v}_b\!\cdot\bm{n}_{ab})
    \big)
    \bigg]
    \notag\\[2pt]
    &\quad
    +\sum_a\sum_{b\neq a}
    \frac{Gm_b}{r_{ab}^{2}}
    \bigg[
    \frac{1}{2}\bm{S}_a\!\cdot\bm{v}_a\times\bm{n}_{ab}
    \bigg(
    v_b^2
    -(\bm{v}_a\!\cdot\bm{v}_b)
    +3(\bm{v}_a\!\cdot\bm{n}_{ab})
    (\bm{v}_b\!\cdot\bm{n}_{ab})
    \bigg)
    \notag\\[2pt]
    &\quad\quad
    +\bm{S}_a\!\cdot\bm{v}_b\times\bm{n}_{ab}
    \big(
    v_b^2
    -(\bm{v}_a\!\cdot\bm{v}_b)
    -3(\bm{v}_a\!\cdot\bm{n}_{ab})
    (\bm{v}_b\!\cdot\bm{n}_{ab})
    \big)
    -\bm{S}_a\!\cdot\bm{v}_a\times\bm{v}_b
    \bigg(
    \frac{1}{2}(\bm{v}_a\!\cdot\bm{n}_{ab})
    +(\bm{v}_b\!\cdot\bm{n}_{ab})
    \bigg)
    \bigg]
    \notag\\[2pt]
    &\quad
    +\sum_a\sum_{b\neq a}
    \frac{G^2m_b}{2r_{ab}^{3}}\,
    \bm{S}_a\!\cdot
    \Big[
    (m_b-2m_a)\bm{v}_a-5m_b\bm{v}_b
    \Big]\times\bm{n}_{ab}
    \notag\\[2pt]
    &\quad
    +\sum_a\sum_{b\neq a}
    \frac{Gm_b}{r_{ab}^{2}}
    S_a^{0i}n_{ab}^i
    \bigg[
    \frac{G(m_a+2m_b)}{r_{ab}}
    -1
    +\frac{3}{2}
    \big(
    (\bm{v}_a\!\cdot\bm{v}_b)
    -v_b^2
    +(\bm{v}_a\!\cdot\bm{n}_{ab})
    (\bm{v}_b\!\cdot\bm{n}_{ab})
    \big)
    \bigg]
    \notag\\[2pt]
    &\quad
    +\sum_a\sum_{b\neq a}
    \frac{3Gm_b}{2r_{ab}^2}
    \Big[
    (\bm{v}_a\!\cdot\bm{n}_{ab})S_a^{0i}v_b^i
    -r_{ab}\,\partial_t S_a^{0i}v_b^i
    \Big]
    \notag\\[2pt]
    &\quad
    +\sum_a\sum_{b\neq a}\sum_{c\neq a,b}
    G^2m_bm_c
    \Bigg\{
    \frac{1}{r_{ab}^2}\,
    \bm{S}_a\!\cdot
    \bigg(
    \frac{\bm{v}_a+2\bm{v}_b-4\bm{v}_c}{r_{bc}}
    +\frac{2(\bm{v}_c-2\bm{v}_b)}{r_{ac}}
    \bigg)\times\bm{n}_{ab}
    \notag\\[2pt]
    &\quad\quad
    +\frac{2}{r_{ab}s_{abc}^{2}}\,
    \bm{S}_a\!\cdot
    \Big[
    4\bm{v}_c-\bm{v}_a-3\bm{v}_b
    -\bm{n}_{ab}
    \big(
    (4\bm{v}_c-\bm{v}_a-3\bm{v}_b)
    \!\cdot\bm{n}_{ab}
    \big)
    \Big]\times(\bm{n}_{ac}+\bm{n}_{bc})
    \notag\\[2pt]
    &\quad\quad
    +\frac{2}{r_{bc}s_{abc}^{2}}\,
    \bm{S}_a\!\cdot
    \Big[\!
    -2\bm{v}_a+\bm{v}_b+\bm{v}_c
    -\bm{n}_{bc}
    \big(
    (-2\bm{v}_a+\bm{v}_b+\bm{v}_c)
    \!\cdot\bm{n}_{bc}
    \big)
    \Big]\times\bm{n}_{ab}
    \notag\\[2pt]
    &\quad\quad
    +\frac{2}{s_{abc}^{3}}
    \Big[
    \bm{v}_a\!\cdot
    (\bm{n}_{ab}-\bm{n}_{ac}-2\bm{n}_{bc})
    +\bm{v}_b\!\cdot
    (3\bm{n}_{ab}+4\bm{n}_{ac}+\bm{n}_{bc})
    \notag\\[2pt]
    &\quad\quad\quad
    -\bm{v}_c\!\cdot
    (4\bm{n}_{ab}+3\bm{n}_{ac}-\bm{n}_{bc})
    \Big]
    \bm{S}_a\!\cdot\bm{n}_{ab}\times
    \left(
    \bm{n}_{ac}+2\bm{n}_{bc}
    \right)
    \notag\\[2pt]
    &\quad\quad
    +\frac{1}{r_{ab}^{2}}
    \left(
    \frac{2}{r_{ac}}+\frac{1}{r_{bc}}
    \right)
    S_a^{0i}n_{ab}^i
    \Bigg\}.
\end{align}
\end{widetext}
The two-body parts of the LO and NLO potentials reproduce the binary covariant-SSC results of Eqs.~(70) and (108) of~\cite{levi2010so}, respectively, after promoting the binary labels to arbitrary body labels and summing over pairs.
The remaining terms are genuine three-body contributions.
They are obtained from the 7 new three-body diagrams (e1)--(e6) and (n2).

\subsection{Canonical Hamiltonian}

As before, we use $L=L_{\rm N}+L_{\rm EIH}-V_{\rm SO}^{\rm LO}$ and obtain the conjugate momentum from $\bm{p}_a=\partial L/\partial \bm{v}_a$ as
\begin{align}
\bm{p}_a
&=
m_a\bm{v}_a
+\frac{m_a v_a^2}{2}\bm{v}_a
\notag\\
&\quad+\sum_{b\neq a}\frac{Gm_a m_b}{r_{ab}}
\left(
3\bm{v}_a
-\frac{7}{2}\bm{v}_b
-\frac{\bm{v}_b\!\cdot\bm{n}_{ab}}{2}\bm{n}_{ab}
\right)
\notag\\
&\quad
-\sum_{b\neq a}\frac{G}{r_{ab}^2}
\left(
m_b\bm{S}_a\times\bm{n}_{ab}
+2m_a\bm{S}_b\times\bm{n}_{ab}
\right).
\end{align}
Inverting this relation to the desired order gives
\begin{align}
\bm{v}_a
&=
\frac{\bm{p}_a}{m_a}
-\frac{p_a^2}{2m_a^3}\bm{p}_a
\notag\\
&\quad+\sum_{b\neq a}\frac{G}{r_{ab}}
\left(
-3\frac{m_b}{m_a}\bm{p}_a
+\frac{7}{2}\bm{p}_b
+\frac{\bm{p}_b\!\cdot\bm{n}_{ab}}{2}\bm{n}_{ab}
\right)
\notag\\
&\quad
+\sum_{b\neq a}
\frac{G}{m_a r_{ab}^2}
\left(
m_b\bm{S}_a\times\bm{n}_{ab}
+2m_a\bm{S}_b\times\bm{n}_{ab}
\right).
\end{align}

We now remove the acceleration terms by a position shift, as in Sec.~\ref{sec:IIIC}, which is equivalent to using the 1PN equations of motion.
Then, we impose the covariant SSC given in Eq.~\eqref{covssc}. To do so, we take its spatial component by setting $\mathsf{a}\to i$ and write the four-velocity in terms of the coordinate velocity, $v^i=u^i/u^0$. Equation~\eqref{covssc} then gives
\begin{align}
    S^{i0}
    &= 
    S^{ij}\frac{e^j_0+e^j_kv^k}{e^0_0+e^0_\ell v^\ell}
    = S^{ij}\bigg(v^j+\frac{1}{2}A^j-2\phi v^j\bigg)
\end{align}
to the desired order, where we used
$e^0_0=1+\phi$, $e^i_0=A^i/2$, $e^0_i=-A_i/2$, and
$e^i_j=(1-\phi)\delta_{ij}$. Substituting the 1PN fields
\begin{align}
    \phi(x) &= -\sum_b \frac{Gm_b}{|\bm{x}-\bm{x}_b|},
    \\
    A^i(x) &=-\sum_b \frac{4Gm_b v_b^i}{|\bm{x}-\bm{x}_b|},
\end{align}
evaluating the result on the worldline of body $a$, and using $S^{0i}=-S^{i0}$, we find
\begin{equation}
    S^{0i}_a
    =
    -
    S^{ij}_a
    \Bigg[
    v_a^j
    +\sum_{b\neq a}\frac{2Gm_b}{r_{ab}}(v_a^j-v_b^j)
    \Bigg].
    \label{sscc}
\end{equation}
This is the relation we use, after writing the velocities in terms of canonical momenta, to eliminate $S^{0i}$ from the Hamiltonian.

We then perform the following noncanonical transformation~\cite{levi2010so}, which maps the covariant SSC to the Pryce--Newton--Wigner SSC and makes the spin variables canonical:
\begin{widetext}
\begin{align}
    \bm{S}_a
    &\to
    \bm{S}_a
    +\frac{p_a^2}{2m_a^2}\bm{S}_a
    -\frac{\bm{p}_a\!\cdot\!\bm{S}_a}{2m_a^2}\bm{p}_a ,
    \\[2pt]
    \bm{x}_a
    &\to
    \bm{x}_a
    +\frac{\bm{S}_a\!\times\bm{p}_a}{2m_a^2}
    \left(
    1-\frac{p_a^2}{4m_a^2}
    \right)
    -\sum_{b\neq a}\frac{Gm_b}{r_{ab}}
    \left(
    \frac{\bm{S}_a\!\times\bm{p}_a}{m_a^2}
    -\frac{3\,\bm{S}_a\!\times\bm{p}_b}{2m_am_b}
    \right).
\end{align}
\end{widetext}
After substituting the 1PN equations of motion, imposing the covariant SSC of Eq.~\eqref{sscc}, and performing the noncanonical transformation above, the Hamiltonian terms contributing to the NLO spin-orbit sector are
\begin{widetext}
\begin{align}
H_{\rm SO}^{\rm NLO}
&=
\sum_a\sum_{b\neq a}
\frac{G}{r_{ab}^2}
\bigg[
\left(\!
-\frac{5m_b\,\bm{p}_a^2}{8m_a^3}
-\frac{\bm{p}_a\!\cdot\bm{p}_b}{4m_a^2}
+\frac{\bm{p}_b^2}{4m_am_b}
-\frac{9(\bm{n}_{ab}\!\cdot\bm{p}_a)(\bm{n}_{ab}\!\cdot\bm{p}_b)}{4m_a^2}
\right)
\bm{S}_a\!\cdot\bm{n}_{ab}\times\bm{p}_a
\notag\\[2pt]
&\quad\quad
+\left(
\frac{\bm{p}_a\!\cdot\bm{p}_b}{m_am_b}
+\frac{3(\bm{n}_{ab}\!\cdot\bm{p}_a)(\bm{n}_{ab}\!\cdot\bm{p}_b)}{m_am_b}
\right)
\bm{S}_a\!\cdot\bm{n}_{ab}\times\bm{p}_b
+\left(
\frac{\bm{n}_{ab}\!\cdot\bm{p}_a}{4m_a^2}
-\frac{\bm{n}_{ab}\!\cdot\bm{p}_b}{m_am_b}
\right)
\bm{S}_a\!\cdot\bm{p}_a\times\bm{p}_b
\bigg]
\notag\\[2pt]
&\quad
+\sum_a\sum_{b\neq a}
\frac{G^2}{r_{ab}^3}
\bigg[\!
-\left(
\frac{11}{2}m_b+\frac{5m_b^2}{m_a}
\right)
\bm{S}_a\!\cdot\bm{n}_{ab}\times\bm{p}_a
+\left(
6m_a+7m_b
\right)
\bm{S}_a\!\cdot\bm{n}_{ab}\times\bm{p}_b
\bigg]
\notag\\[2pt]
&\quad
+\sum_a\sum_{b\neq a}\sum_{c\neq a,b}
G^2
\Bigg\{\!
-\frac{3m_bm_c}{2m_ar_{ac}^2}
\left(
\frac{3}{r_{ab}}+\frac{1}{r_{bc}}
\right)
\bm{S}_a\!\cdot\bm{n}_{ac}\times\bm{p}_a
+\frac{m_c}{r_{ac}^2}
\left(
\frac{5}{2r_{ab}}-\frac{3}{r_{bc}}
\right)
\bm{S}_a\!\cdot\bm{n}_{ac}\times\bm{p}_b
\notag\\[2pt]
&\quad\quad
+\frac{4m_b}{r_{ac}^2}
\left(
\frac{1}{r_{ab}}+\frac{1}{r_{bc}}
\right)
\bm{S}_a\!\cdot\bm{n}_{ac}\times\bm{p}_c
-\frac{m_bm_c}{m_ar_{ab}r_{bc}^2}
\bm{S}_a\!\cdot\bm{n}_{bc}\times\bm{p}_a
\notag\\[2pt]
&\quad\quad
+\frac{3m_b}{2r_{ab}^2r_{ac}}
(\bm{n}_{ac}\!\cdot\bm{p}_c)\,
\bm{S}_a\!\cdot\bm{n}_{ab}\times\bm{n}_{ac}
-\bigg(
\frac{m_b}{r_{ab}^2r_{bc}}
(\bm{n}_{bc}\!\cdot\bm{p}_c)
+\frac{m_bm_c}{m_ar_{ab}r_{bc}^2}
(\bm{n}_{ab}\!\cdot\bm{p}_a)
\bigg)
\bm{S}_a\!\cdot\bm{n}_{ab}\times\bm{n}_{bc}
\notag\\[2pt]
&\quad\quad
+\frac{1}{s_{abc}^2}
\bigg[
\frac{2m_bm_c}{m_ar_{ab}}\,
\bm{S}_a\!\cdot(\bm{n}_{ac}+\bm{n}_{bc})\times\bm{p}_a
+\frac{6m_c}{r_{ab}}\,
\bm{S}_a\!\cdot(\bm{n}_{ac}+\bm{n}_{bc})\times\bm{p}_b
\notag\\[2pt]
&\quad\quad\quad
-\frac{8m_b}{r_{ab}}\,
\bm{S}_a\!\cdot(\bm{n}_{ac}+\bm{n}_{bc})\times\bm{p}_c
+\frac{4m_bm_c}{m_ar_{bc}}\,
\bm{S}_a\!\cdot\bm{n}_{ab}\times\bm{p}_a
-2\,\frac{m_c}{r_{bc}}\,
\bm{S}_a\!\cdot(\bm{n}_{ab}+\bm{n}_{ac})\times\bm{p}_b
\notag\\[2pt]
&\quad\quad\quad
+\bm{S}_a\!\cdot\bm{n}_{ab}\times\bm{n}_{ac}
\bigg(
\frac{2m_bm_c}{m_ar_{ab}}
(\bm{n}_{ab}\!\cdot\bm{p}_a)
+\frac{6m_c}{r_{ab}}
(\bm{n}_{ab}\!\cdot\bm{p}_b)
+\frac{8m_c}{r_{ac}}
(\bm{n}_{ac}\!\cdot\bm{p}_b)
\bigg)
\notag\\[2pt]
&\quad\quad\quad
+2\bm{S}_a\!\cdot\bm{n}_{ab}\times\bm{n}_{bc}
\bigg(
\frac{m_bm_c}{m_ar_{ab}}
(\bm{n}_{ab}\!\cdot\bm{p}_a)
-\frac{2m_bm_c}{m_ar_{bc}}
(\bm{n}_{bc}\!\cdot\bm{p}_a)
\notag\\[2pt]
&\quad\quad\quad\quad
+\frac{3m_c}{r_{ab}}
(\bm{n}_{ab}\!\cdot\bm{p}_b)
+\frac{m_c}{r_{bc}}
(\bm{n}_{bc}\!\cdot\bm{p}_b)
-\frac{4m_b}{r_{ab}}
(\bm{n}_{ab}\!\cdot\bm{p}_c)
+\frac{m_b}{r_{bc}}
(\bm{n}_{bc}\!\cdot\bm{p}_c)
\bigg)
\bigg]
\notag\\[2pt]
&\quad\quad
+\frac{1}{s_{abc}^3}
\bigg[
\bm{S}_a\!\cdot\bm{n}_{ab}\times\bm{n}_{ac}
\bigg(
\frac{4m_bm_c}{m_a}
(\bm{n}_{ab}\!\cdot\bm{p}_a)
-\frac{4m_bm_c}{m_a}
(\bm{n}_{bc}\!\cdot\bm{p}_a)
\notag\\[2pt]
&\quad\quad\quad\quad
+12m_c(\bm{n}_{ab}\!\cdot\bm{p}_b)
+16m_c(\bm{n}_{ac}\!\cdot\bm{p}_b)
+4m_c(\bm{n}_{bc}\!\cdot\bm{p}_b)
\bigg)
\notag\\[2pt]
&\quad\quad\quad
+2\bm{S}_a\!\cdot\bm{n}_{ab}\times\bm{n}_{bc}
\bigg(
\frac{m_bm_c}{m_a}
\left(
2\,\bm{n}_{ab}\!\cdot\bm{p}_a
-2\,\bm{n}_{ac}\!\cdot\bm{p}_a
-4\,\bm{n}_{bc}\!\cdot\bm{p}_a
\right)
\notag\\[2pt]
&\quad\quad\quad\quad
+m_c
\left(
6\,\bm{n}_{ab}\!\cdot\bm{p}_b
+8\,\bm{n}_{ac}\!\cdot\bm{p}_b
+2\,\bm{n}_{bc}\!\cdot\bm{p}_b
\right)
-m_b
\left(
8\,\bm{n}_{ab}\!\cdot\bm{p}_c
+6\,\bm{n}_{ac}\!\cdot\bm{p}_c
-2\,\bm{n}_{bc}\!\cdot\bm{p}_c
\right)
\bigg)
\bigg]
\Bigg\}.
\end{align}
\end{widetext}

The difference between our NLO spin-orbit Hamiltonian and the ADM Hamiltonian of~\cite{hartung2011} is then
\begin{widetext}
\begin{align}
\Delta H
&=
\sum_a\sum_{b\neq a}\frac{Gm_b}{r_{ab}^{2}}\,
\bm{S}_a\!\cdot\!
\Bigg[
\frac{\bm{p}_a\times\bm{n}_{ab}}{m_a}
\Bigg(\!
-\frac{\bm{p}_a\!\cdot\bm{p}_b}{2m_am_b}
+\frac{p_b^2}{2m_b^2}
+\frac{3\,(\bm{p}_a\!\cdot\bm{n}_{ab})(\bm{p}_b\!\cdot\bm{n}_{ab})}{2m_am_b}
-\frac{3\,(\bm{p}_b\!\cdot\bm{n}_{ab})^2}{2m_b^2}
\Bigg)
\notag\\[2pt]
&\quad\quad
-\frac{\bm{p}_a\times\bm{p}_b}{m_am_b}
\Bigg(
\frac{\bm{p}_a\!\cdot\bm{n}_{ab}}{2m_a}
-\frac{\bm{p}_b\!\cdot\bm{n}_{ab}}{m_b}
\Bigg)
\Bigg]
\notag\\[2pt]
&\quad+\sum_a\sum_{b\neq a}\frac{G^2m_b}{2r_{ab}^3}\,
\bm{S}_a\!\cdot\bm{p}_b\times\bm{n}_{ab}
\notag\\[2pt]
&\quad
+\sum_a \sum_{b\neq a}\sum_{c\neq a,b}
G^2\bm{S}_a\!\cdot\!
\Bigg[
\frac{m_c}{2r_{ab}r_{ac}^2}\,\bm{p}_b\times\bm{n}_{ac}
+\frac{\bm{p}_a\times\bm{n}_{ab}}{m_a}
\Bigg(
\frac{m_bm_c}{4r_{bc}^3}
-\frac{m_bm_c\,r_{ab}}{2r_{ac}r_{bc}^3}
-\frac{m_bm_c}{4r_{ab}^2r_{bc}}
\left(
1-\frac{r_{ac}^2}{r_{bc}^2}
\right)
\Bigg)
\notag\\[2pt]
&\quad\quad
+\frac{m_b}{2r_{ab}^2r_{ac}}\,
(\bm{n}_{ac}\!\cdot\bm{p}_c)\,
\bm{n}_{ab}\times\bm{n}_{ac}
\Bigg].
\end{align}
\end{widetext}
The equation
\begin{equation}
    \Delta H=\left\{H,g\right\} = -\frac{dg}{dt}
\end{equation}
is solved by the same generator as in Eq.~\eqref{generator}, with
\begin{equation}
    g_1=-1/2, \quad g_3=1/2,\quad g_2=g_4=g_5=0.
\end{equation} 
This is the same solution found in the binary case in~\cite{levi2010so}.
Thus, this Hamiltonian is also physically equivalent to the ADM Hamiltonian of~\cite{hartung2011}.

\section{Conclusions}
\label{sec:VI}

In this paper, we derived the $N$-body NLO spin-orbit potential and Hamiltonian in the PN-EFT formalism. This extends the known binary PN-EFT results to arbitrary $N$ and gives an independent EFT derivation of the $N$-body Hamiltonian previously obtained in the canonical ADM formalism by Hartung and Steinhoff.

We carried out the calculation in two ways. First, we worked in the generalized canonical gauge, in which the canonical Hamiltonian is obtained directly from the EFT potential. Second, we used a covariant-SSC formulation, followed by a noncanonical transformation to canonical variables. In the generalized canonical gauge, the extension from the binary case to arbitrary $N$ introduces 6 three-body diagrams in the NLO spin-orbit sector, whereas the covariant-SSC derivation introduces 7. In both routes, the resulting canonical Hamiltonian agrees with the $N$-body ADM Hamiltonian of Hartung and Steinhoff after a canonical transformation.

This agreement is a nontrivial check of the ADM result and of our independent PN-EFT derivation. It also shows explicitly how the PN-EFT formalism extends beyond the binary case in the spin-orbit sector. At this order, the only new contributions beyond the binary case are the three-body diagrams; all remaining contributions are obtained by promoting the binary diagrams to arbitrary pairs and summing over body labels. The agreement between the two EFT derivations further confirms the consistency between the generalized canonical gauge and the SSC-based formulation.

This work provides a basis for PN-EFT calculations of spinning gravitating many-body systems.
Natural extensions include the $N$-body NNLO spin-orbit interaction and the $N$-body NLO spin-spin interaction.
Such results would further develop the spin sector beyond binaries and support future GW modeling and data analysis for compact-object systems with more than two bodies.

\acknowledgments
L.W. thanks K.~Imai, K.~Kobayashi, Y.~Murakami, M.~Suzuki and U.~Nguyen for useful discussions on EFT.
This work was supported by the Japan Society for the Promotion of Science (JSPS) Grants-in-Aid for Transformative Research Areas (A) No.~23H04893 and by the Joint Research Program of the Institute for Cosmic Ray Research, The University of Tokyo.
L.W. acknowledges support from the Japanese Government (MEXT) Scholarship.
Some of the calculations were performed with the \texttt{xAct} package~\cite{Garcia} for Mathematica.

%\appendix
%\input{sections/appendix}

\newpage

\bibliographystyle{apsrev4-1}
\bibliography{mybib}

\end{document}